\documentclass[12pt]{article}
\usepackage{epsf,amssymb,psfrag}
\usepackage{ae,aecompl}
\usepackage{graphicx}
\catcode`\@=11
\def \thesection {\arabic{section}.}

\def \be  {\begin{equation}}
\def \ee  {\end{equation}}
\def \ba  {\begin{eqnarray}}
\def \ea  {\end{eqnarray}}
\def \baa {\begin{eqnarray*}}
\def \eaa {\end{eqnarray*}}
\def \bb  {\begin {thebibliography} }
\def \eb  {\end{thebibliography}}
\def \lab #1 {\label{#1}}

\def \matrix #1 {\left(\begin{array}{cc} #1 \end{array}\right)}

\def \tr {\mathop{\rm tr}\nolimits}
\def \Im {\mathop{\rm Im}\nolimits}
\def \Re {\mathop{\rm Re}\nolimits}

\newcommand \widebar [1] {\overline{#1}}

\newcommand \vev [1] {\langle{#1}\rangle}

\newcommand{\as}{\ifmmode\alpha_{\rm s}\else{$\alpha_{\rm s}$}\fi}
\newcommand{\asbar}{\ifmmode\bar{\alpha}_{\rm s}\else{$\bar{\alpha}_{\rm s}$}\fi}
\def \CO {{\cal O}}
\def \CP {{\cal P}}

\def \CM {{\cal M}}

\def \CH {{\cal H}}

\def \CR {{\cal R}}
\def \CD {{\cal D}}
\def \CN {{\cal N}}
\def \CE {{\cal E}}

\font\cmss=cmss12 
\def\inbar{\,\vrule height1.5ex width.4pt depth0pt}
\def\IC{\relax\hbox{$\inbar\kern-.3em{\rm C}$}}
\def\IZ{\relax{\hbox{\cmss Z\kern-.4em Z}}}
\def\IR{{\hbox{{\rm I}\kern-.2em\hbox{\rm R}}}}

\def\IP{{\hbox{{\rm I}\kern-.2em\hbox{\rm P}}}}
\def\II{\hbox{{1}\kern-.25em\hbox{l}}}

\def\numberbysection{\@addtoreset{equation}{section}
                     \def\theequation{\thesection\arabic{equation}}}
\numberbysection

\newcommand \mybf[1] {\mbox{\boldmath$\scriptstyle {#1} $}}
\newcommand \Mybf[1] {\mbox{\boldmath$ {#1} $}}
\newbox\lett\newdimen\lheight\newdimen\lwidth
\def\ontop#1#2{\setbox\lett=\hbox{#2}\lheight\ht\lett
\multiply\lheight by 12 \divide\lheight by 10\relax%
\lwidth\wd\lett \multiply\lwidth by 8 \divide\lwidth by 10\relax #2\kern-\lwidth%
\raise\lheight\hbox{{$\scriptstyle #1$}}\kern.1ex}

\begin{document}

\begin{titlepage}
\begin{flushright}
\begin{tabular}{l}
RUB-TP2-03/04\\
hep-th/0405030
\end{tabular}
\end{flushright}

\vskip3cm
\begin{center}
  {\large \bf   Noncompact $SL(2,\mathbb{R})$ spin chain}

\def\thefootnote{\fnsymbol{footnote}}%
\vspace{1cm}
{\sc M. Kirch},
 \ \ {\sc A.N.~Manashov}\footnote{ Permanent
address:\ Department of Theoretical Physics,  Sankt-Petersburg State University,
St.-Petersburg, Russia}
\\[0.5cm]
\vspace*{0.1cm}
 {\it
Institut f\"ur Theoretische Physik II, Ruhr-Universit\"at Bochum,\\
D-44780 Bochum, Germany}

\vskip2cm

{\bf Abstract:\\[10pt]} \parbox[t]{\textwidth}{
We consider the integrable spin chain model --
the noncompact $SL(2,\mathbb{R})$ spin magnet.
The spin operators are realized as the generators of the
unitary principal series
representation of the $SL(2,\mathbb{R})$ group.
In an explicit form, we construct $\CR-$matrix, the Baxter $\mathbb{Q}-$operator and the
transition kernel to the  representation of the Separated Variables
(SoV). The expressions for the energy and quasimomentum of the
eigenstates in terms of the Baxter $\mathbb{Q}-$operator are derived.
The analytic properties of the eigenvalues of the Baxter operator as a
function of the spectral parameter are established. Applying the
diagrammatic approach, we calculate Sklyanin's integration measure in
the separated variables and obtain the solution to the spectral problem
for the model in terms of the eigenvalues of the $\mathbb{Q}-$operator.
We show that the transition kernel to the SoV representation is
factorized into a product of certain operators each depending on a
single separated variable.}
\vskip1cm

\end{center}

\end{titlepage}

\newpage
\tableofcontents

\setcounter{footnote}{0}

\section{Introduction}
Solutions of many integrable models can be obtained within the
Algebraic Bethe Ansatz (ABA) method~\cite{ABA}.
However, in some cases, this method is not applicable. This happens, in
particular, when the Hilbert space of the model does not possess the
lowest weight vector.
In this case the solution of the model should rely on the methods of
the Baxter $\mathbb{Q}-$operator~\cite{Baxter} and  the
Separated Variables (SoV)~\cite{Sklyanin1,Sklyanin}, which present an alternative
to the ABA approach. The advantage of these two methods is that their
applicability is not subjected to the restrictions of the ABA method, while
the disadvantage is that the
Baxter $\mathbb{Q}-$operator and representation of
the Separated Variables are known only for a limited class of models.
The most well known model of this class is the quantum Toda chain.
The  method of SoV was developed by
Sklyanin~\cite{Sklyanin1} and worked out for the Toda chain by  Kharchev and Lebedev~\cite{KL}.
The Baxter $\mathbb{Q}-$operator for the Toda  chain was constructed
by Pasquier and Gaudin~\cite{PG}.

Recently
the Baxter $\mathbb{Q}-$operator and representation of the
Separated Variables  have been constructed for a number of
models~\cite{KL,PG,SD,Pronko,KSS,FS,DKM-I,DKM,KP,KMS,open}.
Among these models there are so-called noncompact spin magnets.
They are the cousins of the conventional spin magnets like the famous
$XXX$ Heisenberg spin magnet. The difference between these models is that the
spin generators act in different Hilbert spaces. In the case of
the compact magnets it is the finite dimensional space of certain
representations of the $SU(2)$ group. The Hilbert space of  the noncompact magnets
is taken to be the vector space of the unitary representations of the
noncompact group $SL(2,\mathbb{R})$ or
$SL(2,\mathbb{C})$, which are necessarily infinite-dimensional.

Compact $SU(2)$ spin magnets are of special interest in statistical
physics (see e.g. Ref.~\cite{KBI}). The recent interest in
the study of the noncompact magnets originates from high energy
physics (see review Refs.~\cite{LC,BKD}). The studies of the scale
dependence of scattering amplitudes in Quantum Chromodynamics
revealed the surprising fact that the equations governing the scale
dependence of many physically important amplitudes are integrable.
Namely, it has been shown that the Hamiltonian governing the Regge-behaviour
of the hadronic amplitudes is equivalent to the
Hamiltonian of the noncompact $SL(2,\mathbb{C})$ spin magnet~\cite{L,FK}.
Further, it was observed in~\cite{BDM} that the scale dependence of
certain partonic distributions in QCD is governed by Hamiltonians
which are equivalent to those of  the spin chains, both closed and open,
with $SL(2,\mathbb{R})$ symmetry group.
The Hilbert space of the latter is given by the tensor product of the
vector spaces of the discrete series representations of the $SL(2,\mathbb{R})$
group and possesses the lowest weight vector. Therefore the solution
of these models can be obtained
by the ABA method
(see Refs. \cite{BDKM,AB1,DKM99} for the analysis of these
particular models). Contrary, the solution of the $SL(2,\mathbb{C})$
spin magnets relies entirely on the method of the
Baxter $\mathbb{Q}-$operator and Separation of Variables~\cite{DKM-I}.

A general method of constructing the Baxter $\mathbb{Q}-$operator
is not developed yet and the latter is known for a limited class of models only.
For the noncompact $SL(2,\mathbb{R})$
spin chain the Baxter $\mathbb{Q}-$operator was first constructed in Ref.~\cite{SD}.
The approach used in~\cite{SD} is based
on the  Pasquier and Gaudin method~\cite{PG} and suggests an
effective way to resolve the defining equations and obtain the integral kernel of the
$\mathbb{Q}-$operator in an explicit form.
Further, in the case of the noncompact spin magnets it appears to be
quite helpful to interpret the integral
kernels of the operators in question ($\mathbb{Q}-$operator,
transfer matrix, etc.) as Feynman diagrams of a certain type~\cite{DKM-I,DKM}.
Then the analysis of the properties of the models is drastically
simplified and in many cases can be fulfilled diagrammatically.
Moreover, following this approach
one can construct not only the Baxter operator that allows to
determine the energies of the eigenstates, but also the
transition operator to the SoV representation~\cite{DKM-I,DKM,open}
that gives the explicit representation for the eigenfunctions of the
model.
Surprisingly, such a reformulation appears especially effective for the
models where the ABA method does not work, e.g. $SL(2,\mathbb{C})$
spin magnet~\cite{DKM-I}.

In the present paper we consider the
noncompact  spin magnet with the Hilbert space given
by the tensor product of the vector spaces of the unitary principal
series representation of $SL(2,\mathbb{R})$ group. Such models have not been
considered so far and differ drastically in their properties from the
$SL(2,\mathbb{R})$ magnet considered
in the literature.  A particular choice of the Hilbert
space gives this model some similarity with the $SL(2,\mathbb{C})$ spin
magnets. This similarity also appears in the method of the analysis
used. The solution in both cases relies on the Baxter
$\mathbb{Q}-$operator and SoV methods. However, this model possesses
some
specific properties which make it different from both $SL(2,\mathbb{C})$
and $SL(2,\mathbb{R})$ (discrete series) spin magnets. The
two-particle Hamiltonian has both the discrete and continuous
spectrum, the energy of the corresponding eigenstate is not completely fixed by
its two-particle spin and deviates  from the usual $\psi-$function
form. We found that such fundamental object as
$\CR-$operator, Baxter
$\mathbb{Q}-$operator are doubled, i.e. there are two independent
solutions of the Yang-Baxter relation for $\CR-$operator and
two Baxter $\mathbb{Q}-$operators.
The latter satisfy the Baxter equation which entangles them.
Although it is possible to introduce  a linear combination of the Baxter operators
such that each
satisfies the separate equation, they remain entangled through the
quantization conditions.

The paper is organized as follows: In the Sect.~\ref{Prem} we remind
some facts about $SL(2,\mathbb{R})$ group and introduce the notations.
In the Sect.~\ref{R-operator}
we construct the $SL(2,\mathbb{R})$ invariant solution of the Yang-Baxter
relation and discuss its properties. The transfer matrices and
Hamiltonian are built in the Sect.~\ref{TMH}. In the Sect.~\ref{QB}
we construct the Baxter $\mathbb{Q}-$operator and study its properties.
The representation of the  Separated Variables is constructed in the Sect.\ref{SoV}.
The eigenvalues of the Baxter $\mathbb{Q}-$operator can be found in
analytic form for the two-site chain, which is considered in the
Sect.~\ref{N2}. The concluding remarks are given in the Sect.~\ref{Con}
and the Appendix contains some technical details.

\setcounter{equation}{0}
\section{Preliminaries}
\label{Prem}
The aim of this section is to remind some basic facts about the
representations of the group $SL(2,\mathbb{R})$ and introduce the necessary notations.
It is well known  that all
unitary representations of the  group of the real unimodular
matrices $SL(2,\mathbb{R})$ can be organized into three series,
the discrete
ones, the principal and supplementary
continuous series~\cite{Gelfand}.
The latter will not appear in our analysis and therefore is not considered here.

The unitary representation of the principal continuous series is
determined by two numbers, a real $\rho$ and a discrete $\epsilon$,
which takes only two values $0$ and~$1/2$. It is convenient to denote
the pair of numbers $(\alpha,\epsilon)$, where $\alpha$ is  complex and $\epsilon$
is $0$ or $1/2$, by
the bold letter $\Mybf{\alpha}$. We define the sum of two numbers
$\Mybf{\alpha_1}$ and $\Mybf{\alpha_2}$ as follows
\be\label{summi}
\Mybf{\alpha_1}+\Mybf{\alpha_2}=\Mybf{\alpha_3}=(\alpha_1+\alpha_2,\epsilon_3)\,.
\ee
where $\epsilon_3=0$, if $\epsilon_1+\epsilon_2$ is integer, and
$\epsilon_3=1/2$ otherwise. In what follows writing the sum
$(\epsilon_1+\epsilon_2)$ we shall always  imply  such addition rule.
 Next, the zero element is defined as $\Mybf{0}=(0,0)$. By
$-\Mybf{\alpha}$ we shall denote the element inverse
to $\Mybf{\alpha}$,  $\Mybf{\alpha}+(-\Mybf{\alpha})=\Mybf{0}$,
$-\Mybf{\alpha}=(-\alpha,\epsilon)$.
If the first element, $\alpha$,  of the number $\Mybf{\alpha}$ is real, we
shall  call the number $\Mybf{\alpha}$  real as well.
The usefulness of such notations will be clear later.

Thus the unitary representation of the principal continuous series is
labelled by the real~$\Mybf{\rho}$. It
can be realized by the unitary operators $T^{\mybf{\rho}}(g)$ acting on the
Hilbert space $L^2(\mathbb{R})$~\cite{Gelfand}
\be\label{trc}
\left[T^{\mybf{\rho}}(g)\Psi\right](x)~=~
\frac{\sigma_\epsilon(cx+d)}{|cx+d|^{1+2i\rho}}\Psi\left(\frac{ax+b}{cx+d}\right)\,,
\ee
where $g^{-1}=\left(\begin{array}{cc}a&b\\c&d\end{array}\right)$ and
the sign factor is
$\sigma_\epsilon(x)=[\textrm{sign}(x)]^{2\epsilon}$.
All representations  $T^{\mybf{\rho}}$ (except of the representation
$T^{(0,1/2)}$ )
are irreducible. The two
representations  $T^{\mybf{\rho}}$ and  $T^{\mybf{\rho'}}$ are
not equivalent provided that  $\Mybf{\rho}\neq-\Mybf{\rho'}$. The unitary
operator
$\CM_{\mybf{\rho}}$ which intertwines the representations
$T^{\mybf{\rho}}$ and  $T^{-\mybf{\rho}}$
($T^{-\mybf{\rho}}\CM_{\mybf{\rho}}=
\CM_{\mybf{\rho}}T^{\mybf{\rho}}$)
is defined uniquely up to a phase factor~\cite{Gelfand}
\be\label{un-map}
\left[\CM_{\mybf{\rho}}\Psi\right](x)~=~\left(\sqrt{\pi}\,
A(\Mybf{\gamma})\right)^{-1}\,
\int_{-\infty}^\infty dy \,\Psi(y)\,\frac{\sigma_\epsilon(x-y)}{|x-y|^{2(1-s)}}\,,
\ee
where the parameter $s$
(conformal spin) is determined as $s\equiv 1/2+i\rho$ and
$\Mybf{\gamma}=(2-2s,\epsilon)$. The function $A(\Mybf{\alpha})$ is
given by the following expression
\be\label{A}
A(\Mybf{\alpha})~=~e^{-i\pi\epsilon}\,\left(\frac{\alpha}{2}\right)^{2\epsilon}\,
a(\epsilon+\alpha/2)\,,
\ee
where
$
a(x)~=~\Gamma(1/2-x)/\Gamma(x)\,.
$
This function, which appears naturally in the  course of the
calculations,
satisfies the following conditions
\be\label{AA}
A(\Mybf{\alpha})\,A(\Mybf{1}-\Mybf{\alpha})~=~(-1)^{2\epsilon}\,,\ \ \
\ \ \ A(\Mybf{\alpha})^*~=~(-1)^{2\epsilon}\,A(\Mybf{\alpha^*})\,,
\ee
where $\Mybf{\alpha}^*=(\alpha^*,\epsilon)$.
The normalization factor chosen in (\ref{un-map})
ensures that the intertwining operator satisfies the following equation
 $\CM_\rho^\dagger=\CM_{-\rho}$~\footnote{Henceforth we shall
not display explicitly the label
$\Mybf{\rho}$ of the operator $\CM$ since it completely fixed by the
transformation properties of the function the operator $\CM$ acts on.}.

The three generators of the $s\ell(2)$  algebra corresponding to the
realization of the representation $T^{\mybf{\rho}}$~(\ref{trc}) have the form
\be\label{generators}
S_-=-\partial_x,\ \ \ S_+=x^2\partial_x+2s x, \ \ \  S_0=x\partial_x+s\,.
\ee
These operators  are antihermitean and obey the standard
$s\ell(2)$ commutation relations
\be\label{com-rel}
[S_0,S_\pm]=\pm S_\pm,\ \ \ \ \ \ \ [S_+,S_-]=2S_0\,.
\ee

The tensor product of two representations of the principal unitary
series can be decomposed into the direct integral of the
representations of the same type and the direct sum of the
representations of the discrete series, $\CD_{h}^\pm$,~\cite{Pk}
\begin{equation}\label{Dec}
T^{\mybf{\rho}_1}\otimes T^{\mybf{\rho}_2}=2\int_{0}^{\infty} d\rho\,
T^{\mybf{\rho}}\,+\sum^{\infty}_{h=1+(\epsilon_{1}+\epsilon_{2})/2}(\CD_{h}^+\oplus
\CD_{h}^-)\,.
\end{equation}
The representation of the continuous series, $T^{\mybf{\rho}}$,
enters into the direct integral with the  multiplicity two.
The operators separating the irreducible components and the other
necessary details can be found in the Appendix~\ref{AppA}.

We conclude this section with the following remark. Let us divide all
tensor products $T^{\mybf{\rho_1}}\otimes T^{\mybf{\rho_2}}$ in two
subsets
depending on the value, $0$ or $1/2$, of the sum $\epsilon_1+\epsilon_2$.
Then it is easy to see that all representations inside each group are
unitary equivalent to each other. In particular, the representation of
the first group,  $\epsilon_1+\epsilon_2=0$, are equivalent to $T^{(0,0)}\otimes T^{(0,0)}$,
and those in the second group, $\epsilon_1+\epsilon_2=1/2$, are equivalent
to  $T^{(0,0)}\otimes T^{(0,1/2)}$
\footnote{The representation $T^{(0,1/2)}$ is reducible and equivalent
  to the $\CD^+_{1/2}\oplus\CD^{-}_{1/2}$~\cite{Gelfand}.}.
To show this let us consider  the operators $V(\Mybf{\alpha})$,
\be\label{V-op}
\left[V(\Mybf{\alpha})\Psi \right](x_1,x_2)~=~
\frac{\sigma_\epsilon(x_1-x_2)\,}{|x_1-x_2|^{2i\alpha}}\,\Psi(x_1,x_2)\,,
\ee
and $U(\Mybf{\alpha})$,
\be\label{U-op}
U(\Mybf{\alpha})~=~\left(\CM\otimes
\mathbb{I}\right)\,V(\Mybf{\alpha})\,
\left(\CM\otimes \mathbb{I}\right) \,,
\ee
where $\Mybf{\alpha}=(\alpha,\epsilon)$ and
the operator $\CM$ defined in~(\ref{un-map}) intertwines the
representations  $T^{\mybf{\rho_1}}$ and  $T^{-\mybf{\rho_1}}$. It is
obvious that for real $\Mybf{\alpha}$, the operators $V(\Mybf{\alpha})$
and  $U(\Mybf{\alpha})$  are unitary and that they intertwine the
representation $T^{\mybf{\rho}_1}\otimes T^{\mybf{\rho}_2}$ with
$T^{\mybf{\rho}_1+\mybf{\alpha}}\otimes T^{\mybf{\rho}_2+\mybf{\alpha}}
$ and with $T^{\mybf{\rho}_1-\mybf{\alpha}}
\otimes T^{\mybf{\rho}_2+\mybf{\alpha}} $,
respectively.
Therefore the combination, $U(\Mybf{\alpha})\,V(\Mybf{\beta})$, with suitably chosen
$\Mybf{\alpha},\Mybf{\beta}$ intertwines any two representations inside
each group. We notice also that for a real $\Mybf{\alpha}$,
\be\label{UV}
V(\Mybf{\alpha})^\dagger=V(-\Mybf{\alpha}), \ \ \ \ \ \
U(\Mybf{\alpha})^\dagger=U(-\Mybf{\alpha})\,.
\ee

\section{${\cal R}-$operator}
\label{R-operator}
In this section we shall construct the solution of the Yang-Baxter
relation
\be\label{YB-1}
\CR_{12}(\Mybf{u})\,\CR_{13}(\Mybf{u}+\Mybf{v})\,\CR_{23}(\Mybf{v})~=~
\CR_{23}(\Mybf{v})\,\CR_{13}(\Mybf{u}+\Mybf{v})\,\CR_{12}(\Mybf{u})\,.
\ee
The operator $\CR_{ik}$ in the above equation  acts on the tensor product of the
spaces $V_i\otimes V_k$ and depends on the spectral parameter
$\Mybf{u}=(u,\epsilon)$, i.e. $\CR(\Mybf{u})=\CR(u,\epsilon)$.
Each space $V_i$ is equivalent to the Hilbert
space $L^2(\mathbb{R})$ and carries the representation
of the principal continuous series, $T^{\mybf{\rho}_i}$,
of the $SL(2,\mathbb{R})$ group.
Let us notice that  Eq.~(\ref{YB-1})
contains two equations involving operators $\widetilde R(u)=\CR(u,0)$ and
$\widehat R(u)=\CR(u,1/2)$. Indeed, having put $\Mybf{u}=(u,0)$ and
$\Mybf{v}=(v,0)$ in  Eq.~(\ref{YB-1}) and taking into account that
$\Mybf{u}+\Mybf{v}=(u+v,0)$,
one gets the Yang-Baxter equation for
the operators $\widetilde R$. At the same time, choosing $\Mybf{u}=(u,1/2)$ and
$\Mybf{v}=(v,1/2)$ one finds that the argument of the operator
$\CR_{13}$ remains the same,
 $\Mybf{u}+\Mybf{v}=(u+v,0)$, and  therefore, in this case, Eq.~(\ref{YB-1})
involves two $\widehat R$ operators and one $\widetilde R(u+v)$.
At a given choice of the spectral parameter these two equations are
contained in one equation~(\ref{YB-1}).
It will be seen later that such, slightly unusual, choice of the spectral
parameter is quite natural for this model.

\begin{figure}[t]
\psfrag{a}[cc][cc]{\small{$\Mybf{\alpha}$}}
\psfrag{b}[cc][cc]{\small{$\Mybf{\beta}$}}
\psfrag{g}[cc][cc]{\small{$\Mybf{\gamma}$}}
\psfrag{x}[cc][cc]{$x$}
\psfrag{y}[cc][cc]{$y$}
\psfrag{eq}[cc][cc]{$=\sqrt{\pi}
A(\Mybf{\alpha})A(\Mybf{\beta})/A(\Mybf{\gamma})$}
\centerline{\epsfxsize14.0cm\epsfbox{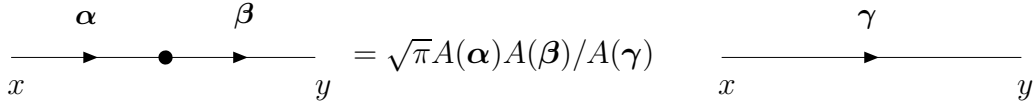}} \vspace*{0.5cm}
\caption{The diagrammatic representation of the chain relation,
  Eq.~(\ref{chain}).
The black dot denotes the integration vertex and
$\Mybf{\gamma}=\Mybf{\alpha+\beta-1}$.}
\label{chain-diag}
\end{figure}
\begin{figure}[t]
\psfrag{a}[cc][cc]{\small{$\Mybf{\alpha}$}}
\psfrag{b}[cc][cc]{\small{$\Mybf{\beta}$}}
\psfrag{g}[cc][cc]{\small{$\Mybf{\gamma}$}}
\psfrag{ap}[cc][cc]{\small{$\Mybf{\alpha}'$}}
\psfrag{bp}[cc][cc]{\small{$\Mybf{\beta}'$}}
\psfrag{gp}[cc][cc]{\small{$\Mybf{\gamma}'$}}
\psfrag{x}[cc][cc]{$x$}
\psfrag{y}[cc][cc]{$y$}
\psfrag{z}[cc][cc]{$z$}
\psfrag{eq}[cl][cc]{$=
\sqrt{\pi}
A(\Mybf{\alpha},\Mybf{\beta},\Mybf{\gamma})$}
\centerline{\epsfxsize12.0cm\epsfbox{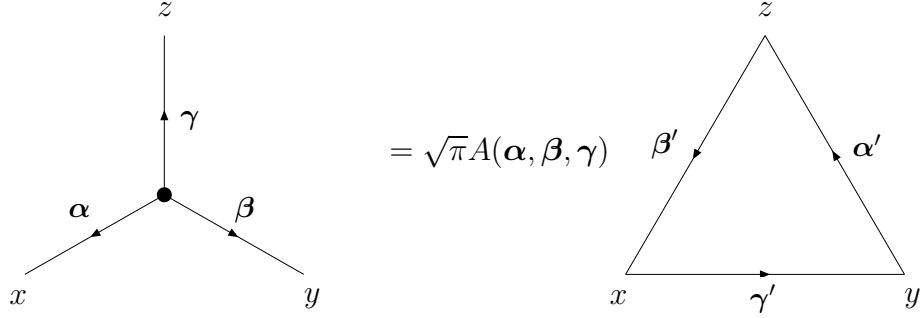}} \vspace*{0.5cm}
\caption{The diagrammatic representation of the star-triangle relation
 (Eq.~(\ref{star-triangle})).
The primed indices are defined as $\Mybf{x}'=1-\Mybf{x}$.}
\label{st-diag}
\end{figure}
We shall look for the solution of~(\ref{YB-1})
in the form of the integral operator
\be
\left[\CR_{12}(\Mybf{u})\Psi \right](x_1,x_2)~=~\int dy_1\,dy_2\,
R_{\mybf{u}}(x_1,x_2|y_1,y_2)\,
\Psi(y_1,y_2)\,,
\ee
and impose the additional restrictions of the $SL(2,\mathbb{R})$
invariance.
The relation~(\ref{YB-1}) leads to certain equations on the integral
kernel. We shall propose
an ansatz for the kernel $R_{\mybf{u}}(x_1,x_2|y_1,y_2)$, which is
motivated by the form of the kernel of the $\CR$ operator for the
$SL(2,\mathbb{C})$ magnet~\cite{DKM-I},
and show that  it
satisfies  Eq.~(\ref{YB-1}). To write down  the kernel
in a compact form let us define the function (propagator)
$D_{\mybf{\alpha}}(x)$,
\be\label{propagator}
 D_{\mybf{\alpha}}(x)=\frac{\sigma_\epsilon(x)}{|x|^{\alpha}}\,,
\ee
which depends on real $x$ and has the index
$\Mybf{\alpha}=(\alpha,\epsilon)$.~
The kernel of
the $\CR-$matrix is given by the product of four propagators
\be\label{R-anz}
R_{\mybf{u}}(x_1,x_2|y_1,y_2)~=~r_{\mybf{\rho}_1
\mybf{\rho}_2}(\Mybf{u})
\,D_{\mybf{\alpha}_4}(y_2-x_1)\,D_{\mybf{\alpha}_3}(x_1-x_2)
\,D_{\mybf{\alpha}_2}(x_2-y_1)\,D_{\mybf{\alpha}_1}(y_1-y_2)\,,
\ee
where $r_{\mybf{\rho}_1 \mybf{\rho}_2}(\Mybf{u})$ is the normalization
coefficient to be defined later and $\Mybf{u}=(u,\epsilon)$ is the
spectral parameter. The requirement of the $SL(2,\mathbb{R})$
invariance of the kernel imposes the following constraints on the indices
$\Mybf{\alpha}_i$
\ba\label{constraints}
\Mybf{\alpha_1}+\Mybf{\alpha_2}=(2-2s_1,\epsilon_1)\,,&\ \ \ \ \ &
\Mybf{\alpha_3}+\Mybf{\alpha_4}=(2s_1,\epsilon_1)\,,\nonumber \\
\Mybf{\alpha_1}+\Mybf{\alpha_4}=(2-2s_2,\epsilon_2)\,,&\ \ \ \ \ &
\Mybf{\alpha_2}+\Mybf{\alpha_3}=(2s_2,\epsilon_2)\,.
\ea
Therefore it is sufficient to fix only one index to restore all others.
Again, using similarity with the $SL(2,\mathbb{C})$ magnet (see Ref.~\cite{DKM-I}) we put
\be\label{alpha4}
\Mybf{\alpha}_4=(1+i\rho_1-i\rho_2-iu,\epsilon)\,
\ee
that gives rises to the following values of the other indices
\ba\label{indices}
\Mybf{\alpha}_1&=&(-i\rho_1-i\rho_2+iu,\epsilon_2+\epsilon)\,,\nonumber\\
\Mybf{\alpha}_2&=&(1-i\rho_1+i\rho_2-iu,\epsilon_1+\epsilon_2+\epsilon)\,,\\
\Mybf{\alpha}_3&=&(i\rho_1+i\rho_2+iu,\epsilon_1+\epsilon)\nonumber\,.
\ea
The proof of the Yang-Baxter relation is based on the integral identity
for  
propagators -- the so-called
star-triangle relation. It is  well known in the perturbative Quantum
Field Theory and is widely used for the evaluation of the multi-loop
Feynman diagrams~\cite{VPK}. As will be seen later, it is quite natural
to represent the integral kernels  of the operators under consideration
in the form of Feynman diagrams. It allows one to carry out an analysis of
the properties of the model diagrammatically, which in many cases results
in  considerable   simplifications. Therefore, below we collect some
elements of the "diagram technique" we use in our analysis,

\begin{figure}[t]
\psfrag{a1}[cc][cc]{$\Mybf{\alpha}_1$}
\psfrag{a2}[cc][cc]{$\Mybf{\alpha}_2$}
\psfrag{a3}[cc][cc]{$\Mybf{\alpha}_3$}
\psfrag{a4}[cc][cc]{$\Mybf{\alpha}_4$}
\psfrag{x1}[cc][cc]{$x_1$}
\psfrag{x2}[cc][cc]{$x_2$}
\psfrag{y1}[cc][cc]{$y_1$}
\psfrag{y2}[cc][cc]{$y_2$}
\psfrag{a}[cc][cc]{$\Mybf{\alpha}$}
\psfrag{b}[ll][cc]{$=\quad D_{\mybf{\alpha}}(x_1-x_2)$}
\centerline{\epsfxsize12.0cm\epsfbox{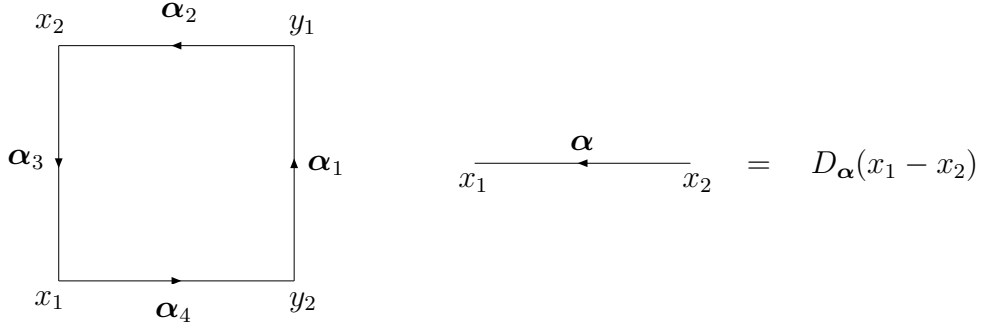}} \vspace*{0.5cm}
\caption{Diagrammatical representation of the kernel of the $\CR$-operator,
 Eq.~(\ref{R-anz}).
}\label{Ruf}
\end{figure}
\begin{itemize}
\item Fourier transform
\be\label{momentum}
\int dx\,e^{ipx}\,\CD_{\mybf{\alpha}}(x)~=~
\sqrt{\pi}\,2^{1-\alpha}\,
A(\Mybf{\alpha})\,\CD_{\mybf{1-\alpha}}(p),
\ee
where $\Mybf{1}=(1,0)$ and the function $A(\Mybf{\alpha})$
($\Mybf{\alpha}=(\alpha,\epsilon))$ is defined in Eq.~(\ref{A}).
\item {Chain relation}
\be\label{chain}
\int dy\,\CD_{\mybf{\alpha}}(x-y)\CD_{\mybf{\beta}}(y-z)=
\sqrt{\pi}
\frac{A(\Mybf{\alpha})A(\Mybf{\beta})}{A(\Mybf{\gamma})}\,\CD_{\mybf{\gamma}}(x-z),
\ee
where $\Mybf{\gamma}=\Mybf{\alpha}+\Mybf{\beta}-\Mybf{1}$.
\item {Star-triangle relation}
\ba\label{star-triangle}
&&\int dw\,\CD_{\mybf{\alpha}}(x-w)\CD_{\mybf{\beta}}(y-w)\CD_{\mybf{\gamma}}(z-w)=\\[2mm]
&&\ \ \ \ \ \ \ \ \ \ =\sqrt{\pi}\,
A(\Mybf{\alpha},\Mybf{\beta},\Mybf{\gamma})\,\CD_{\mybf{1-\alpha}}(z-y)
\CD_{\mybf{1-\beta}}(x-z)\CD_{\mybf{1-\gamma}}(y-x).\nonumber
\ea
Here $A(\Mybf{\alpha},\Mybf{\beta},\Mybf{\gamma})\equiv
A(\Mybf{\alpha}) A(\Mybf{\beta}) A(\Mybf{\gamma})$
and indices $\Mybf{\alpha},\Mybf{\beta},\Mybf{\gamma}$
satisfy the {\it uniqueness} condition
 $\Mybf{\alpha}+\Mybf{\beta}+\Mybf{\gamma}=(2,0)$.
\end{itemize}
The diagrammatic representation of the above identities is given in Figs.~\ref{chain-diag} and
\ref{st-diag}. There the arrow directed from point $y$ to $x$
and labelled by the index $\Mybf{\alpha}$ denotes the
propagator $\CD_{\mybf{\alpha}}(x-y)$ and the black dot is used for the
integration vertex.

At last, we give the following representation for the delta function
\be\label{delta}
\delta(x)~=~\lim_{\alpha\rightarrow 0}\frac{a(\alpha/2)}{\sqrt{\pi}}\,
\frac{1\phantom{|x|}}{|x|^{1-\alpha}}.
\ee
which can be obtained from  Eq.~(\ref{momentum}).


\begin{figure}[t]
\psfrag{1p}[cc][cc]{$1'$}
\psfrag{2p}[cc][cc]{$2'$}
\psfrag{3p}[cc][cc]{$3'$}
\psfrag{3}[cc][cc]{$3$}
\psfrag{2}[cc][cc]{$2$}
\psfrag{1}[ct][ct]{$1$}
\centerline{\epsfxsize15.0cm\epsfbox{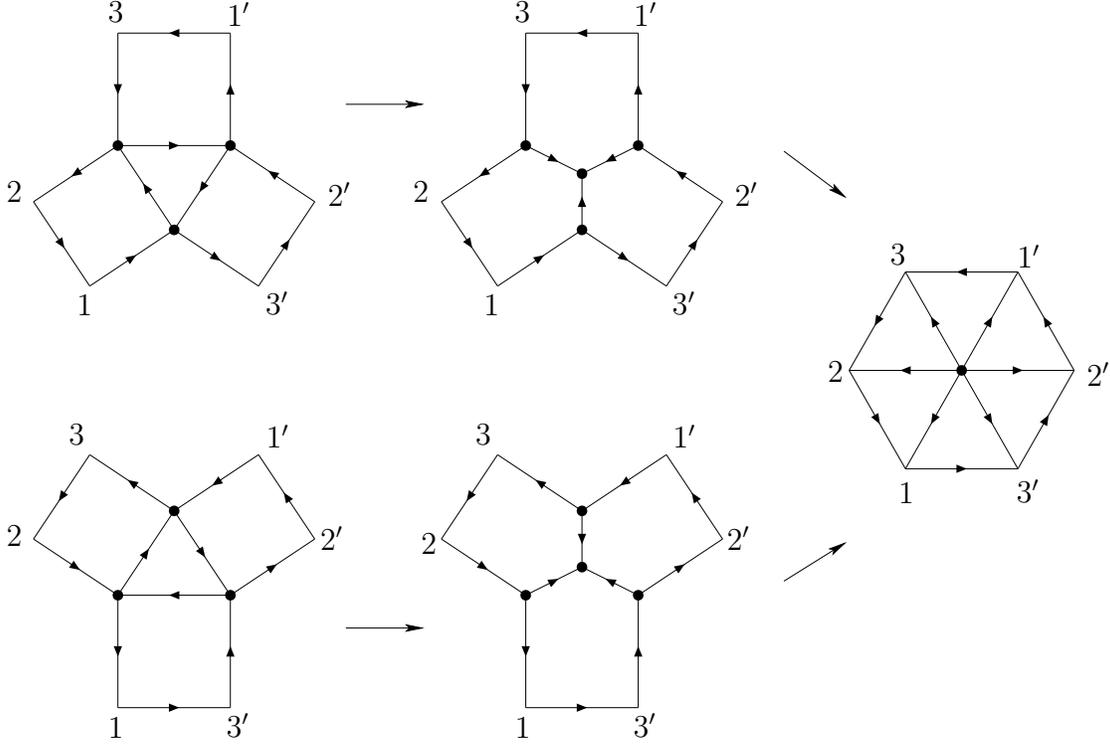}} \vspace*{0.5cm}
\caption{The diagrammatic proof of the Yang-Baxter relation,
Eq.~(\ref{YB-1}).
}\label{YBf}
\end{figure}
\vskip 0.5cm

The proof of the Yang-Baxter relation for the $\CR-$operator
defined in Eq.~(\ref{R-anz}) can be carried out diagrammatically.
To this end let us notice   that graphically the $\CR$ operator is
represented by the box diagram~(see Fig.~\ref{Ruf}).
Then the lhs and rhs of the Yang-Baxter relation~(\ref{YB-1})
are represented by the diagrams shown in the the Fig.~\ref{YBf}, left
up and lower diagrams, respectively. To prove the Yang-Baxter equation one has to
show that these diagrams are equal.
The proof repeats the one given in Ref.~\cite{DKM-I} and
is based on the use of the star-triangle relation~(\ref{star-triangle}).
Indeed, an examination of the indices of the central triangles in those diagrams  shows that
they satisfy the   \textit{uniqueness} condition and, therefore,
can be replaced by the star diagrams~(\ref{star-triangle}) (see second
up and lower diagrams in Fig.~\ref{YBf}).
The new triple vertices in each diagram appear to be unique as
well. In the next step one replaces the corresponding star subdiagrams by triangles.
The resulting diagram, in both cases, is the hexagon diagram.
Restoring all factors which arise in the course of above transformations,
see Eq.~(\ref{star-triangle}), one finds that the final diagrams coincide.

One can also show that $\CR-$operator satisfies the Yang-Baxter
relation involving two Lax operators
\be\label{Lax-YB}
L_1(u)\,L_2(v+u)\,\CR_{12}(\Mybf{u})~=~\CR_{12}(\Mybf{u})\,L_{2}(v)\,L_1(u)\,,
\ee
where, as usual, $L_1(u)=L(u)\otimes \mathbb{I}$,
$L_2(u)=\mathbb{I}\otimes L(u)$.
The Lax operator is defined in the
standard way
\be\label{Lax}
L(u)~=~u+i\left(\begin{array}{cc}
  S_0 & S_- \\
  S_+ & -S_0
\end{array}\right)
\ee
and is used for the construction of the auxiliary transfer matrix.

Let us note that the above Yang-Baxter relation holds for the both values
of the $\epsilon$, which enters the spectral parameter of the
$\CR-$operator, i.e. Eq.~(\ref{Lax-YB}) holds both for $\CR_{12}(u,0)$
and $\CR_{12}(u,1/2)$.

\subsection{Properties of $\CR-$operator}
\begin{figure}[t]
\psfrag{a}[cc][cc]{$(a)$}
\psfrag{b}[cc][cc]{$(b)$}
\psfrag{x1}[cc][cc]{$x_1$}
\psfrag{x2}[cc][cc]{$x_2$}
\psfrag{y1}[cc][cc]{$y_1$}
\psfrag{y2}[cc][cc]{$y_2$}
\centerline{\epsfxsize13.0cm\epsfbox{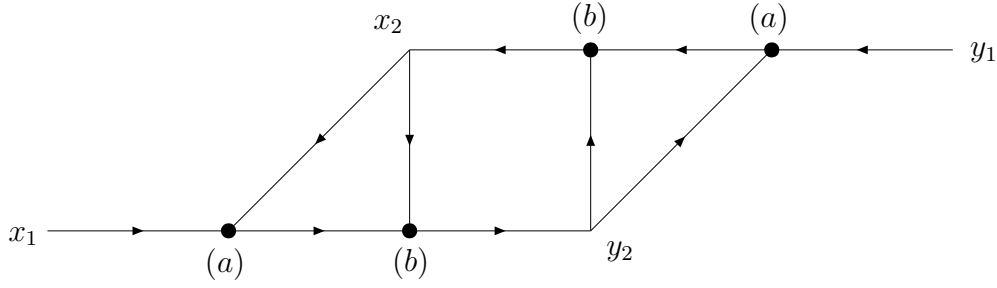}}
\vspace*{0.5cm}
\caption{The diagrammatic representation of the rhs of Eqs.~(\ref{defRp}).}
\label{URU}
\end{figure}

Since the tensor  products $T^{\mybf{\rho}_1}\otimes T^{\mybf{\rho}_2}$
fall into two classes of equivalence, it is natural to expect that the
$\CR$ operators should be unitary equivalent as well.
Indeed, it
follows directly from the definition of the $\CR$ operator ~(\ref{R-anz})
that
$\CR_{\mybf{\rho}_1+\mybf{\alpha},\mybf{\rho}_2+\mybf{\alpha}}\sim
  V(\Mybf{\alpha})\, \CR_{\mybf{\rho}_1\mybf{\rho}_2}\, V(-\Mybf{\alpha})$.
In the general case the following relation holds:
\be\label{defRp}
\CR_{\mybf{\rho}_1\mybf{\rho}_2}(\Mybf{u})=
U(\Mybf{\alpha})V(\Mybf{\beta})\,
\CR_{(0,0),(0,\epsilon_{1}+\epsilon_{2})}(u,0)\,V(-\Mybf{\beta})\,U(-\Mybf{\alpha})\,,
\ee
where $\Mybf{\alpha}=((\rho_2-\rho_1)/2,\epsilon)$ and
$\Mybf{\beta}=((\rho_2+\rho_1)/2,\epsilon_1+\epsilon)$.
Here some comments are in order. The diagram corresponding to
the integral kernel of the rhs of Eq.~(\ref{defRp})
is drawn in Fig.~\ref{URU}.  The central box corresponds to the
kernel of the operator
$V(\Mybf{\beta})\CR_{(0,0),(0,\epsilon_{1}+\epsilon_{2})}(u,0)V(-\Mybf{\beta})$,
and the star subdiagrams from the left and right sides represent the
kernel of the $U$ operators.
The straightforward application of
the star-triangle relation~(\ref{star-triangle}) in a way indicated in
Fig.~\ref{URU}~--~one applies the star-triangle relation to the vertices $(a)$
and then to the vertices $(b)$~--~turns the
diagram shown in Fig.~\ref{URU} into the box diagram  with indices
given by~(\ref{alpha4}), (\ref{indices}). Therefore, up to some
numerical factor, the resulting diagram coincides with those for the
$\CR-$operator.

Let us remind that the normalization factor, $r_{\mybf{\rho}_1\mybf{\rho}_2}(\Mybf{u})$,
in Eq.~(\ref{R-anz}) is so far arbitrary. Since an arbitrary
$\CR-$operator is related by Eq.~(\ref{defRp}) to the operator
$\CR_{(0,0),(0,0)}({u},0)$ or
$\CR_{(0,0),(0,1/2)}({u},0)$, we fix the normalization of the latter,
and for all other operators take Eq.~(\ref{defRp}) for a
definition.
Namely, we fix the normalization factor in Eq.~(\ref{R-anz}) for
the two selected operators as follows
\be\label{ru}
r_{(0,0),(0,\epsilon_2)}(u,0)~=~
{2^{-2iu}}\frac{e^{i\pi\epsilon_2}}{\pi}\,A(iu,0)\,A(iu,\epsilon_2)\,,
\ee
where the function $A(\Mybf{\alpha})$ is introduced in (\ref{A}). Then one finds that the
definition of the $\CR-$operator via Eq.~(\ref{defRp})
is equivalent to the following choice of
the normalization factor in Eq.~(\ref{R-anz})
\begin{equation}\label{rn}
r_{\mybf{\rho}_1\mybf{\rho}_2}(\Mybf{u})~=~\frac1\pi\,e^{i\pi(\epsilon_1+\epsilon_2)}
\,{2^{-2iu}}\,
A(i\rho_2-i\rho_1+iu,\epsilon)\,A(i\rho_1-i\rho_2+iu,\epsilon+\epsilon_1+\epsilon_2)\,.
\end{equation}
We remind that the sum $(\epsilon_1+\epsilon_2)$ is defined by
Eq.~(\ref{summi}).
Such a choice of normalization ensures that the
$\CR-$operator is unitary for the real
$\Mybf{u}$,
\be\label{R-unitary}
\CR_{12}(\Mybf{u})\CR_{12}^\dagger(\Mybf{u})=\mathbb{I}\,.
\ee
Using Eq.~(\ref{delta}) it can be easily checked that for $\Mybf{\rho}_1=\Mybf{\rho}_2$
\be\label{R0}
\CR_{12}(\Mybf{0})~=~(-1)^{2\epsilon_1}\,P_{12}\,,
\ee
where $P_{12}$ is the permutation operator, $[P_{12}\Psi](x_1,x_2)=\Psi(x_2,x_1)$.

Moreover, taking into account that
$V(-\Mybf{\beta})
U(-\Mybf{\alpha})=(U(\Mybf{\alpha})V(\Mybf{\beta}))^\dagger$, (see Eq.~~(\ref{UV})),
one concludes that Eq.~(\ref{defRp}) establishes the unitary
equivalence between the arbitrary $\CR-$operator and  the operator
$\CR_{(0,0),(0,0)}(\Mybf{u})$ (or
 $\CR_{(0,0),(0,1/2)}(\Mybf{u})$).  Thus in order to determine the
spectrum of an arbitrary operator $\CR_{\mybf{\rho}_1\mybf{\rho}_2}$ it is
sufficient to calculate the spectrum of the $\CR$ operator for the special
values $\Mybf{\rho}_1=(0,0)$ and $\Mybf{\rho}_2=(0,\epsilon_2)$.
The eigenfunctions of the $\CR$ operator are fixed by the
$SL(2,\mathbb{R})$ invariance of the latter. To find them one
has to construct the operators separating the irreducible components in
the tensor product of  two representations of the continuous
series,~Eq.~(\ref{Dec}). The tensor product decomposition~(\ref{Dec})
contains the representations both of the discrete and continuous
series. The operators ($\Pi^{h,\pm}_{\mybf{\rho}_1\mybf{\rho}_2}(x_1,x_2,w)$)
separating the representations of the discrete
series, $\CD_h^\pm$, are given in Eq. (\ref{Prod})~\footnote{We
remind that $h$ takes integer or half-integer values.}.
Next, since the
representations of the continuous series enter the tensor product
decomposition ~Eq.~(\ref{Dec}) with double multiplicity, there is arbitrariness in the
choice of the projectors.
We defined the projectors to the continuous series,
$\Pi^{\mybf{\rho},\varepsilon}_{\mybf{\rho}_1\mybf{\rho}_2}(x_1,x_2,y)$,~\footnote{The
parameter $\varepsilon=0,1/2$ marks different projectors.} (see
Eq.~(\ref{Proc}))
 in such
way that they
possess definite parity with respect to the interchange of the arguments $x_1$ and $x_2$.
A direct check shows that namely such combinations diagonalizes
the $\CR-$operator. Introducing the notations for
eigenvalues of the $\CR-$operator on the corresponding eigenfunctions
\ba\label{eigen-c}
\CR_{\mybf{\rho}_1\mybf{\rho}_2}(u,\epsilon)\,
\left( \Pi^{\mybf{\rho},\varepsilon}_{\mybf{\rho}_1\mybf{\rho}_2} \right)^*&=&
R_{\rho,\varepsilon}(u,\epsilon)
\left( \Pi^{\mybf{\rho},\varepsilon}_{\mybf{\rho}_1,\mybf{\rho}_2} \right)^*\,,\\[2mm]
\label{eigen-d}
\CR_{\mybf{\rho}_1\mybf{\rho}_2}(u,\epsilon)\,
\left( \Pi^{h,\pm}_{\mybf{\rho}_1\mybf{\rho}_2} \right)^*&=&
R_{h}^{\pm}(u,\epsilon)
\left( \Pi^{h,\pm}_{\mybf{\rho}_1,\mybf{\rho}_2} \right)^*\,,
\ea
one obtains after some calculations
\begin{itemize}
\item $\epsilon_1=\epsilon_2$
\ba\label{ei-ee}
R_{\rho,\varepsilon}(u,\epsilon)&=&\frac1\pi\,
\left[\sin(i\pi u)+(-1)^{2\epsilon+2\varepsilon}\sin(\pi s)\right]\Gamma(1-s-iu)\Gamma(s-iu)\,,
\\[2mm]
R_{h}^{\pm}(u,\epsilon)&=&\,(-1)^h\frac{\Gamma(h-iu)}{\Gamma(h+iu)}\,,
\ea
\item $\epsilon_1\neq\epsilon_2$
\ba\label{ei-ene}
R_{\rho,\varepsilon}(u,\epsilon)&=&\frac1\pi\,
\left[\cos(i\pi u)+(-1)^{2\epsilon+2\varepsilon}\cos(\pi s)\right]\Gamma(1-s-iu)\Gamma(s-iu)\,,
\\
R_{h}^{\pm}(u,\epsilon)&=&\,(-1)^{h-1/2}\frac{\Gamma(h-iu)}{\Gamma(h+iu)}\,,
\ea
where $s=1/2+i\rho$.
\end{itemize}

\subsection{Transfer matrices and Hamiltonians}
\label{TMH}
\begin{figure}[t]
\psfrag{x1}[cc][cc]{$x_1$}
\psfrag{x2}[cc][cc]{$x_2$}
\psfrag{xn}[cc][cc]{$x_N$}
\psfrag{y0}[cc][cc]{$y_0$}
\psfrag{y1}[cc][cc]{$y_1$}
\psfrag{y2}[cc][cc]{$y_2$}
\psfrag{yn}[cc][cc]{$y_N$}
\psfrag{a1}[cc][cc][1][-45]{\small{$\Mybf{\alpha}_1$}}
\psfrag{a2}[cc][cc][1][45]{\small{$\Mybf{\alpha}_2$}}
\psfrag{a3}[cc][cc][1][-45]{\small{$\Mybf{\alpha}_3$}}
\psfrag{a4}[cc][cc][1][45]{\small{$\Mybf{\alpha}_4$}}
\psfrag{dots}[cc][cc]{\large{$\cdots\quad$}}
\centerline{\epsfxsize15cm\epsfbox{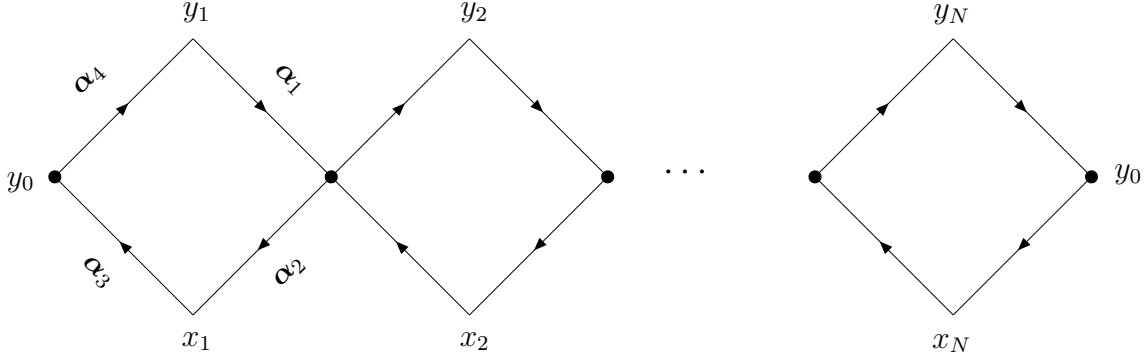}}\vspace*{0.5cm}
\caption{The diagrammatic representation of the kernel of the transfer
  matrix,
Eq. (\ref{TM}). The indices $\Mybf{\alpha}_i$ of the lines on the $k$-th box are given
  by Eqs. (\ref{alpha4}) and (\ref{indices}) with $\Mybf{\rho}_1=\Mybf{\rho}_0$
and $\Mybf{\rho}_2=\Mybf{\rho}_k$.}\label{TMf}
\end{figure}

\begin{figure}[t]
\centerline{\epsfxsize10.0cm\epsfbox{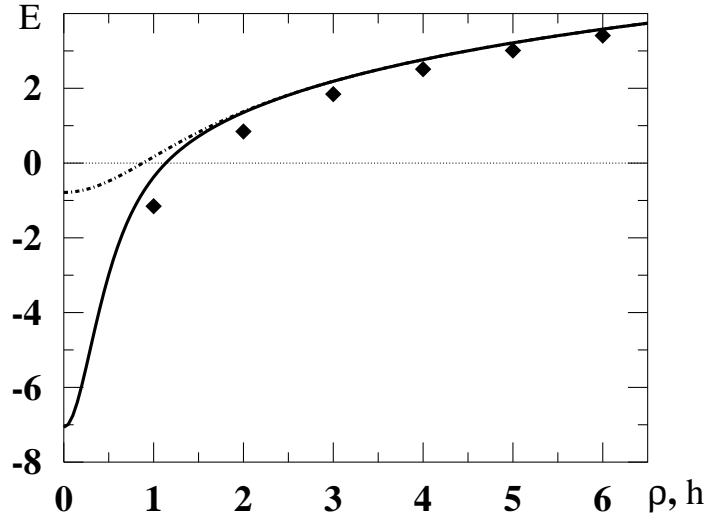}} \vspace*{0.5cm}
\caption{The dispersion curves for the two-particle Hamiltonian.
The curves for $E^{(0)}(\rho)$ and $E^{(1/2)}(\rho)$ are shown by  the
solid line and the  dot-dashed line, respectively.
The discrete eigenvalues $E^\pm(h)$\,($h=1,2,..$) are depicted by the black diamonds.
}\label{disp}
\end{figure}

Having obtained the  solutions of  the Yang-Baxter relation~(\ref{YB-1}),
one can  construct an infinite set of the commuting $SL(2,\mathbb{R})$ invariant
operators (transfer matrices)
\be\label{TM}
\mathbb{T}_{\mybf{\rho}_0}(\Mybf{u})~=~
\tr_{\mybf{\rho}_0}\left(
\CR_{\mybf{\rho}_0\mybf{\rho}_1}(\Mybf{u})\ldots\CR_{\mybf{\rho}_0\mybf{\rho}_N}(\Mybf{u})
\right)\,,
\ee
where the trace is taken over the auxiliary space
$V^{\mybf{\rho}_0}$. They depend on the spectral parameter $\Mybf{u}$
and the spin of the auxiliary space, $\Mybf{\rho}_0$.
The diagrammatic representation of the transfer matrix is shown in Fig.~\ref{TMf}.
Invoking the standard arguments~\cite{QISM}, one finds that the
transfer matrices form the family of the mutually commuting operators
\be\label{T-comm}
\left[\mathbb{T}_{\mybf{\rho}_0}(\Mybf{u}),\mathbb{T}_{\mybf{\rho}^\prime_0}(\Mybf{v})
\right]~=~0\,.
\ee
In the case of the homogeneous chain, (when all quantum spaces carry
the same representation of the $SL(2,\mathbb{R})$ group), i.e. when
$\Mybf{\rho}_1=\Mybf{\rho}_2=\ldots=\Mybf{\rho}_N\equiv\Mybf{\rho}_q$,
it is possible to construct the operator (Hamiltonian) which has
two-particle structure. Indeed, choosing $\Mybf{\rho}_0=\Mybf{\rho}_q$
and taking into account the property~(\ref{R0}), one obtains
\be\label{H}
\CH_N=i\left[\frac{d\phantom{u}}{du}\,\ln\,\mathbb{T}_{\mybf{\rho}_q}(\Mybf{u})\right]
\Biggl |_{\mybf{u}=(0,0)}~=~H_{12}+\cdots+H_{N-1,N}+H_{N,1}\,,
\ee
where the two-particle Hamiltonian $H_{k,k+1}$ is given by
\be\label{H12}
H_{12}~=~i\frac{d\phantom{u}}{du}
\ln \CR_{\mybf{\rho}_1\mybf{\rho}_2}(\Mybf{u})
\biggl |_{\mybf{u}=\mybf{0}}\,.
\ee
Introducing the  notation $E^{\varepsilon}(\rho)$ for the eigenvalues of the pairwise Hamiltonian
corresponding to the eigenfunctions
$\Pi^{\mybf{\rho},\varepsilon}_{\mybf{\rho}_1\mybf{\rho}_2}$~\footnote{
Since $\Mybf{\rho}_1=\Mybf{\rho}_2$, one gets that $\Mybf{\rho}=(\rho,0)$.}
and  $E^{\pm}(h)$ for those corresponding to the eigenfunctions
$\Pi^{h,\pm}_{\mybf{\rho}_1\mybf{\rho}_2}$,
one derives from Eqs.~(\ref{ei-ee}) the following expressions for the energies
\ba\label{ei-H}
E^{\varepsilon}(\rho)&=&\psi\left(\frac12+i\rho
\right)~+~\psi\left(\frac12-i\rho
\right)-(-1)^{2\varepsilon}\frac{\pi}{\cosh\pi\rho}\,,\\[2mm]
E^{\pm}(h)&=&2\psi(h)\,.
\ea
Herein $\psi(x)$, as usual, denotes the logarithmic derivative of the Euler $\Gamma-$function.
Thus one sees that the energies corresponding to the discrete levels
are double-degenerate. The separation between  the continuous branches
being large in region of small $\rho$, vanishes rapidly for large
$\rho$. The corresponding dispersion curves are shown in
Fig.~\ref{disp}. Note also, that although the eigenvalues of the
two-particle Hamiltonian do not depend on the $\Mybf{\rho}_q$,
this dependence reveals itself  in the
eigenfunctions.  For the chains  with the number of sites $N>2$
the energies will explicitly depend on the values of $\Mybf{\rho}_q$.

Let us consider now the
simplest among the transfer matrix --- the auxiliary transfer
matrix. It is given by the trace of the product of the Lax
operators~(\ref{Lax})
\be\label{a-t}
t_N(u)~=~\tr\left(L_1(u)\ldots L_N(u)\right).
\ee
The auxiliary transfer matrix, $t_N(u)$, is the polynomial of degree
$N$ in $u$ with operator valued coefficients
\be\label{t-q}
t_N(u)=2u^N+\sum_{k=2}^N u^{N-k} q_k\,.
\ee
The operators $q_k$ (integrals of motion) are differential operators
in $x_k$. In particular,
\be\label{q2}
q_2~=~-{\vec{S}}^2~-~N(\rho_q^2+1/4)\,,
\ee
where $\vec{S}=\vec{S}_1+\ldots+\vec{S}_N$ is the operator of the total
spin. The possible values of the total Casimir operator, ${\vec{S}}^2$, are restricted
by the $SL(2,\mathbb{R})$ invariance, namely,
$\vec{S}^2=h(h-1)$ for the representations of the discrete series and
$\vec{S}^2=-1/4-\rho^2$ for the representations of the continuous
series.

Further,
by virtue of Eq.~(\ref{Lax-YB}) one derives that
\be\label{t-comm}
\left[\vec{S},t_N(u)\right]~=~
\left[t_N(u),t_N(v)\right]~=~\left[t_N(u),\mathbb{T}_{\mybf{\rho}_0}(\Mybf{v}))\right]~=~
\left[t_N(u),\CH_N\right]~=~0\,.
\ee
It follows from (\ref{t-comm}) that
the auxiliary transfer matrix and Hamiltonian share a common
set of eigenfunctions. Thus the eigenfunctions of the Hamiltonian
can be labelled by the eigenvalues of the integrals of motion
$q_2,\ldots,q_N$ and one of the $SL(2,\mathbb{R})$ generators, which is usually
chosen  to be $S_-$. However, in the case under consideration, these
quantum numbers do not fix the eigenfunction completely. Indeed,
as seen already for the two-site spin chain, the eigenfunctions of the
Hamiltonian, corresponding to $E^{0}(\rho)$ and $E^{1/2}(\rho)$  (see
Eq.~(\ref{ei-ee})) have the same quantum numbers. Therefore, the
spectrum of the transfer matrix is double degenerate, and fixing of the
total momentum, $iS_-=p$, does not remove this degeneracy. Nevertheless,
the number of eigenfunctions with identical integrals of motion, $q_k$, is
finite. One can always resolve the degeneracy on this finite
subspace by fixing an additional quantum number, for example the value of
the quasimomentum $\theta$. We remind that $e^{i\theta}$ is the
eigenvalue of the operator of the cyclic permutation
\be\label{P}
\CP\Psi(x_1,x_2,\ldots,x_N)=\Psi(x_2,\ldots,x_N,x_1)\,,
\ee
which commutes with both auxiliary transfer matrix and Hamiltonian,
$[\CP,t_N(u)]=[\CP,\CH_N]=0$. Thus to uniquely determine the eigenstate,
one should the specify the total momentum, $p$, and the following set
of the quantum numbers, $\mathbf{q}=\{\theta,q_2,\ldots,q_N\}$.

\section{Baxter $\mathbb{Q}-$operator}
\label{QB}
The solution of the eigenvalue problem for the Hamiltonian~(\ref{H})
can be obtained within the method of the Baxter
$\mathbb{Q}-$operators~\cite{Baxter}. The standard method of solving spin chain
models, the Algebraic Bethe Ansatz (ABA)  method~\cite{ABA,QISM}, is
not applicable in the case under consideration since the Hilbert space
of the model does not have  a normalizable lowest weight vector.

We remind that by definition the Baxter $\mathbb{Q}-$operator is the
operator which acts on the Hilbert space of the model, depends on the
spectral parameter and satisfies the following requirements
\begin{itemize}
\item Commutativity
\be\label{comm-Q}
\left[\mathbb{Q}(\Mybf{u}),\mathbb{Q}(\Mybf{v})\right]~=~
\left[\mathbb{Q}(\Mybf{u}),t_N({v})\right]~=~\left[
\mathbb{Q}(\Mybf{u}),\CH_N\right]~=~0\,,
\ee
\item Baxter equation~\footnote{Similarly to the Yang-Baxter equation~(\ref{YB-1})
  this equation contains two finite-difference equations which
  entangle the Baxter operators $\mathbb{Q}(u,0)$ and $\mathbb{Q}(u,1/2)$.}
\be\label{BE}
t_N(u)\,\mathbb{Q}(\Mybf{u})~=~(u+is_q)^N\,\mathbb{Q}(\Mybf{u}+\mathbf{i})~+~(u-is_q)^N
\,\mathbb{Q}(\Mybf{u}-\mathbf{i})\,.
\ee
\end{itemize}
In the case under consideration, the spectral parameter, $\Mybf{u}$,
as natural to expect,
has the same nature as the spectral parameter of the $\CR$ matrix, $\Mybf{u}=(u,\epsilon)$.
We also used the notation $\mathbf{i}=(i,1/2)$ and $s_q=1/2+i\rho_q$ in
Eq.~(\ref{BE}).


As usual we look for the operator
$\mathbb{Q}$ in the form of an integral operator. Then Eq.~(\ref{BE})
results in a particular differential equation on the kernel of the
operator $\mathbb{Q}$. One can find the general solution of this
equation using approach developed in~\cite{PG,SD}. This is based on the
invariance of the auxiliary transfer matrix~(\ref{a-t}) with respect to local
rotations of the Lax operators
\be\label{Lax-rot}
L_k(u)\to {\tilde L}_k(u)~=~M_k^{-1}\,L_k(u)M_{k+1}\,,
\ee
where $M_k$ are arbitrary $2\times 2$ nondegenerate matrices.
Then one can show that the function $Y_{\mybf{u}}(\vec{x},\vec{y})$,
\be\label{Y}
Y_{\mybf{u}}(\vec{x},\vec{y})~=~\prod_{k=1}^{N}D_{\mybf{\alpha}_u}(x_k-y_{k+1})\,
D_{\mybf{\beta}_u}(x_k-y_{k})\,,
\ee
where
\ba\label{alphabeta}
\Mybf{\alpha}_u&=&(s_q-iu,\epsilon+\epsilon_q)\,, \ \ \ \ \ \ \
\Mybf{\beta}_u=(s_q+iu,\epsilon)\,, \ \ \ \ \Mybf{\alpha}_u+\Mybf{\beta}_u=(2s_q,\epsilon_q)\,,
\ea
 satisfies the following
equation
\be\label{YEq}
t_N(u)\,Y_{\mybf{u}}(\vec{x},\vec{y})~=~(u+is)^N\,Y_{\mybf{u}+\mathbf{i}}(\vec{x},\vec{y})+
(u-is)^N\,Y_{\mybf{u}-\mathbf{i}}(\vec{x},\vec{y})
\ee
at arbitrary values of variables $y_1,\ldots,y_N$.
This means that the convolution of the function
$Y_{\mybf{u}}(\vec{x},\vec{y})$ with an arbitrary function of
$Z(\vec{y})$ also satisfies Eq.~(\ref{BE}).
\begin{figure}[t]
\psfrag{a}[cc][cc]{\scriptsize{$\Mybf{\alpha}$}}
\psfrag{b}[cc][cc]{\scriptsize{$\Mybf{\beta}$}}
\psfrag{g}[cc][cc]{\scriptsize{$\Mybf{\gamma}$}}
\psfrag{am}[cc][cc]{\scriptsize{$\Mybf{\alpha}'$}}
\psfrag{bm}[cc][cc]{\scriptsize{$\Mybf{\beta}'$}}
\psfrag{gm}[cc][cc]{\scriptsize{$\Mybf{\gamma}'$}}
\psfrag{x1}[cc][cc]{$x_1$}
\psfrag{x2}[cc][cc]{$x_2$}
\psfrag{x3}[cc][cc]{$x_3$}
\psfrag{xn}[cc][cc]{$x_N$}
\psfrag{y1}[cc][cc]{$y_1$}
\psfrag{y2}[cc][cc]{$y_2$}
\psfrag{yn}[cc][cc]{$y_N$}
\psfrag{eq}[cc][cc]{\footnotesize{$=\left(\sqrt{\pi}A(\Mybf{\alpha},
\Mybf{\beta},\Mybf{\gamma})\right)^N$}}
\psfrag{dots}[cc][cc]{$\cdots$}
\centerline{\epsfxsize17.0cm\epsfbox{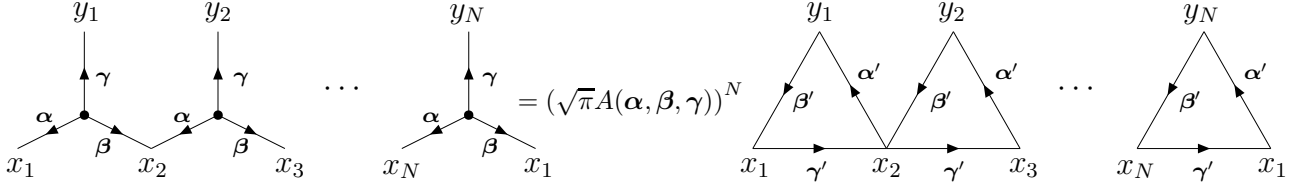}} \vspace*{0.5cm}
\caption{The kernel of the $\mathbb{Q}$-operator in two equivalent
  representations.
The indices are explained in the text;
  $\Mybf{x}^\prime=\Mybf{1-x}$.}
\label{Q-diag}
\end{figure}
The details of the derivation of Eq.(\ref{YEq})
can be found in Ref.~\cite{DKM-I} where the
similar case of the $SL(2,\mathbb{C})$ spin chain was considered.
Let us only note here that one can consider the function
$Y_{\mybf{u}}(\vec{x},\vec{y})$ as an array of two functions which
depends on the complex parameter $u$ only,
$Y_{\mybf{u}}=\left(Y_{(u,0)}, Y_{(u,1/2)}\right)$. Then Eq.~(\ref{YEq})
takes form of the matrix finite-difference equation.

Thus the kernel of the Baxter operator can be written in the form
\be\label{QZ}
Q_{\mybf{u}}(\vec{x},{\vec{x}}^\prime)~=~\int \prod_{k=1}^N dy_k
\,Y_{\mybf{u}}(\vec{x},\vec{y})\, Z(\vec{y},{\vec{x} }^\prime)\,,
\ee
where the function $ Z(\vec{y},\vec{x}') $ does not depend on the
spectral parameter $\Mybf{u}$. The restrictions on this function can
be deduced from the requirement of the commutativity of the Baxter
$\mathbb{Q}-$operator and the auxiliary transfer matrix.  To this end
let us note that the function $Y_{\mybf{u}}(\vec{x},\vec{y})$
possesses the following property
\be\label{YY}
Y_{\mybf{u}}(\vec{x},\vec{y})~=~(-1)^{2N\epsilon_q}\, \CP \,Y_{\mybf{u}^\prime}(\vec{y},\vec{x}),
\ee
where $\Mybf{u}^\prime=(-u,\epsilon+\epsilon_q)$ and $\CP$ is the
operator of the cyclic permutation~(\ref{P}). Then taking into account
Eq.~(\ref{YEq}) and the invariance of the transfer matrix under
cyclic permutations, $[\CP,t_N(u)]=0$, one derives
\be\label{xy}
t_N(u,\vec{S}(\rho_q,x))\,Y_{\mybf{u}}(\vec{x},\vec{y})~=~(-1)^N\,t_N(-u,\vec{S}(\rho_q,y))\,
Y_{\mybf{u}}(\vec{x},\vec{y})~
=~\,t_N(u,-\vec{S}(\rho_q,y))\,Y_{\mybf{u}}(\vec{x},\vec{y})\,.
\ee
Here we indicated explicitly that the transfer matrix $t_N(u)$ is
expressed in terms of the differential operators~(\ref{generators})
acting on $x$ or $y$-coordinates. Then, noticing that the integration
by parts results in the  change $\rho_q\to-\rho_q$ in the generators $\int dy
\left[S(\rho_q,y) \Phi(y)\right] \Psi(y)=\int dy
\Phi(y)\left[S(-\rho_q,y) \Psi(y)\right]$ and taking advantage of Eq.~(\ref{xy}), one derives that the requirements $\left[t_N(u),\mathbb{Q}(\Mybf{u})\right]=0$
gives the following equation on the function $Z(\vec{x},\vec{y})$
\be\label{Zt}
t_N(u,S(-\rho_q,y))\,Z(\vec{y},\vec{x}')~=~Z(\vec{y},\vec{x}')\,t_N(u,S(\rho_q,x'))\,.
\ee
The simplest solution to this equation corresponds to the operator $Z$
intertwining the generators $\vec{S}_k(-\rho_q)$ and
$\vec{S}_k(\rho_q)$. Thus one can choose the operator $Z$ to be proportional to
the product $\CM_1\otimes \CM_2\otimes\ldots\otimes\CM_N$,
where the intertwining operators $\CM_k$ are defined in~(\ref{un-map}).
For later convenience we put
\be\label{Z}
Z(\vec{x},\vec{y})~=~\prod_{k=1}^N D_{\mybf{\gamma}}(y_{k-1}-x_k)\,.
\ee
with $\Mybf{\gamma}=(2-2s_q,\epsilon_q)$.

As usual, we visualize
the kernel of the $\mathbb{Q}-$operator as a Feynman diagram, which is shown
in the lhs of Fig.~\ref{Q-diag}. In the rhs of Fig.~\ref{Q-diag}, we
give the equivalent representation of the kernel~(\ref{QZ}) which can
be obtained from the diagram on the lhs with the help of the star-triangle
relation~(\ref{star-triangle}). The graphical representation of the
kernel is very convenient for the analysis of the properties of the
Baxter  $\mathbb{Q}-$operator. As an example we prove the first
relation in~(\ref{comm-Q}), namely
$\left[\mathbb{Q}(\Mybf{u}),\mathbb{Q}(\Mybf{v})\right]=0$.

\begin{figure}[t]
\psfrag{av}[cc][cc]{\scriptsize{$\Mybf{\alpha}_v$}}
\psfrag{bv}[cc][cc]{\scriptsize{$\Mybf{\beta}_v$}}
\psfrag{aup}[cc][cc]{\scriptsize{$\Mybf{1-\alpha}_u$}}
\psfrag{bup}[cc][cc]{\scriptsize{$\Mybf{1-\beta}_u$}}
\psfrag{x1}[cc][cc]{$x_1$}
\psfrag{x2}[cc][cc]{$x_2$}
\psfrag{x3}[cc][cc]{$x_3$}
\psfrag{x4}[cc][cc]{$x_4$}
\psfrag{xn}[cc][cc]{$x_N$}
\psfrag{y1}[cc][cc]{$y_1$}
\psfrag{y2}[cc][cc]{$y_2$}
\psfrag{yn}[cc][cc]{$y_N$}
\psfrag{dots}[cc][cc]{$\cdots$}
\centerline{\epsfxsize15.0cm\epsfbox{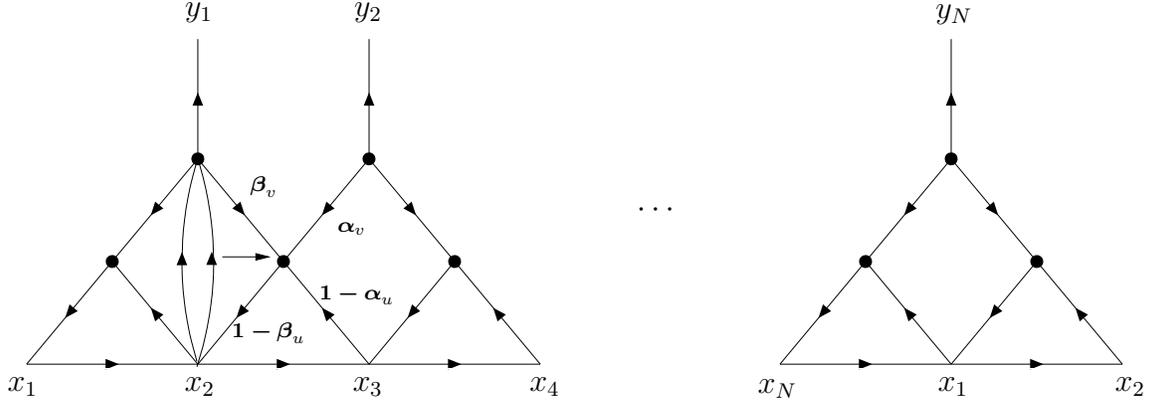}} \vspace*{0.5cm}
\caption{The diagrammatic representation of the product
$\mathbb{Q}(u)\mathbb{Q}(v)$.}\label{QQ-diag}
\end{figure}
The diagrammatic representation of the integral kernel of the operator
$\mathbb{Q}(\Mybf{u})\,\mathbb{Q}(\Mybf{v}) $ is shown in Fig.~\ref{QQ-diag}.
There we represent the kernel of the operator $\mathbb{Q}(\Mybf{v})$ by
the diagram shown in the lhs of  Fig.~\ref{Q-diag}, while for the
operator
$\mathbb{Q}(\Mybf{u})$ we choose the alternative representation given
by the diagram in
the rhs of  Fig.~\ref{Q-diag}.
We also inserted two propagators with opposite indices in the central rhombus in
Fig.~\ref{QQ-diag}.
Since $D_{\mybf{\sigma}}(x)\,D_{-\mybf{\sigma}}(x)=1$,
such an insertion does not alter  the value of the diagram.
\begin{figure}[t]\vspace{.5cm}
\psfrag{ab}[cc][cc][1][90]{\small{$\Mybf{\beta}_{u}-\Mybf{\beta}_{v}$}}
\psfrag{ar}[cc][cc][1][90]{\small{$\Mybf{\beta}_{u}-\Mybf{\beta}_{v}$}}
\psfrag{av}[cc][cc]{\small{$\Mybf{\alpha}_v$}}
\psfrag{bv}[cc][cc]{\small{$\Mybf{\beta}_{v}$}}
\psfrag{mau}[cc][cc]{\small{$\Mybf{1}-\Mybf{\alpha}_{u}$}}
\psfrag{mbu}[cc][cc]{\small{$\Mybf{1}-\Mybf{\beta}_u$}}
\psfrag{mav}[cc][cc]{\small{$\Mybf{1}-\Mybf{\alpha}_{v}$}}
\psfrag{au}[cc][cc]{\small{$\Mybf{\alpha}_u$}}
\psfrag{mbv}[cc][cc]{\small{$\Mybf{1}-\Mybf{\beta}_v$}}
\psfrag{bu}[cc][cc]{\small{$\Mybf{\beta}_{u}$}}
\psfrag{eq}[cc][cc]{$=$}
\psfrag{la}[cc][cc]{$A(\Mybf{\alpha}_u,\Mybf{\beta}_{u},)$}
\psfrag{ra}[cc][cc]{$A(\Mybf{\alpha}_{v},\Mybf{\beta}_{v})$}
\centerline{\epsfxsize15.0cm\epsfbox{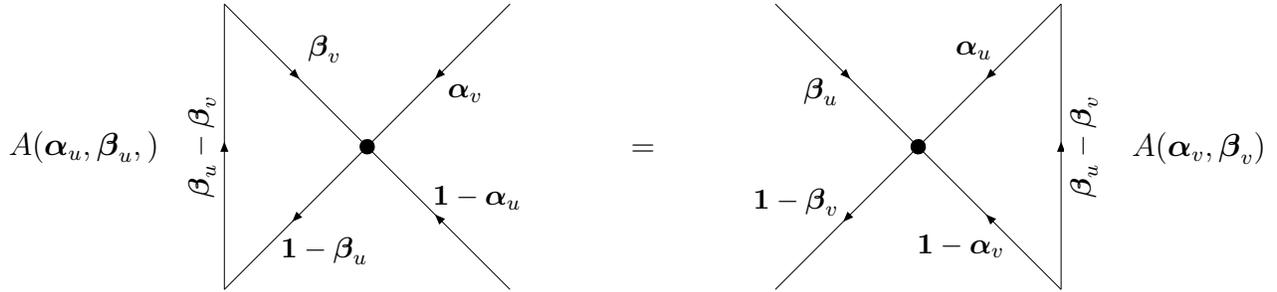}}\vspace*{0.5cm}
\caption{The diagrammatic representation of the exchange relation.
The indices take the special values
$\Mybf{\alpha}_x=(s+ix,\epsilon_x+\epsilon)$ and
$\Mybf{\beta}_x=(s-ix,\epsilon)$,
where $s=1/2+i\rho$, ($\Mybf{\alpha}_u+\Mybf{\beta}_u=
\Mybf{\alpha}_v+\Mybf{\beta}_v=(2s,\epsilon)$).}\label{exchange-diag}
\end{figure}

The proof is based on the use of an additional diagrammatic identity--
exchange relation--which is shown in  Fig.~\ref{exchange-diag}. This
identity can be easily derived with the help of the star-triangle
relation~(\ref{star-triangle}). Further,
to prove the commutativity of the Baxter operators we choose the index
of the inserted propagator to be,
$\Mybf{\sigma}=\Mybf{\beta}_u-\Mybf{\beta}_v$. Then one can use the
exchange relation to move this inserted propagator, $D_{\mybf{\sigma}}$, to the right.
The displacing  of the propagator $D_{\mybf{\sigma}}$ from left
to right of the cross subdiagram alters the indices of the
propagators and produces some scalar factor  in the way shown in Fig.~\ref{exchange-diag}.
After
repeating this operation $N$ times the propagator resumes its initial
position and annihilates the propagator $D_{-\mybf{\sigma}}$.
Therefore the  resulting diagram will have  the same form as that in
Fig.~\ref{QQ-diag}. Examing the indices of the propagators and the
scalar factor one finds that they correspond to the diagram for the operator
$\mathbb{Q}(\Mybf{v})\,\mathbb{Q}(\Mybf{u})$, if the representation in the lhs of Fig.~\ref{Q-diag} is used for $\mathbb{Q}(\Mybf{u})$ and for operator $\mathbb{Q}(\Mybf{v})$ those in the rhs of the same figure.
Thus one concludes
that
\be\label{QQ-QQ}
\mathbb{Q}(\Mybf{u})\,\mathbb{Q}(\Mybf{v})~=~\mathbb{Q}(\Mybf{v})\,\mathbb{Q}(\Mybf{u})\,.
\ee
The Eqs.~(\ref{Y}), (\ref{Z}) together with (\ref{QZ}) define the
Baxter $\mathbb{Q}-$operator which
satisfies  Eqs.~(\ref{comm-Q}), (\ref{BE}). (Strictly speaking we did
not show yet that the Baxter operator commutes with the Hamiltonian but will do it later on.)

\subsection{Properties of $\mathbb{Q}-$operator}
Let us examine the Baxter $\mathbb{Q}-$operator
for the special values of the spectral parameter such that indices
$\Mybf{\alpha}$ or $\Mybf{\beta}$ become equal zero.
Using the diagrammatic representation of the latter given in Fig.~\ref{Q-diag},
one finds that for $\Mybf{u}=(is_q,0)$ ($\Mybf{\beta}=\Mybf{0}$)
the lines with the index $\Mybf{\beta}$ disappear and the integral over centers
of the star diagrams can be easily calculated with help of Eq.~(\ref{chain}). A short calculation gives
\be\label{beta0}
\mathbb{Q}(is_q,0)~=~c(\Mybf{\rho}_q)^N\,\mathbb{I}\,,
\ee
where $\mathbb{I}$ is the unit operator and the normalization constant
is given by
\be\label{cN}
c(\Mybf{\rho})~=~\pi |A(1+2i\rho,\epsilon)|^2=\frac{\pi}{\rho}\tanh^{4\epsilon-1}(\pi\rho)\,.
\ee
Repeating this calculation for $\Mybf{u}=(-is_q,\epsilon_q)$
($\Mybf{\alpha}=\Mybf{0}$), one finds
\be\label{alpha0}
\mathbb{Q}(-is_q,\epsilon_q)~=~c(\Mybf{\rho}_q)^N\,\CP\,,
\ee
where $\CP$ is the operator of the cyclic
permutations~(\ref{P}). Therefore, the operator $\CP$ can be expressed
as the ratio of the Baxter operator at two special points
\be\label{P-Q}
\CP~=~\mathbb{Q}(-is_q,\epsilon_q)/\mathbb{Q}(is_q,0)\,.
\ee

Next, using the diagrammatic representation for the Baxter
$\mathbb{Q}-$operator~
(Fig.~\ref{Q-diag})
and for the transfer matrix~(Fig.~\ref{TMf}) one finds after
some algebra
\ba\label{T-QQ}
T_{\mybf{\rho}_0}(\Mybf{u})&=&\chi(\Mybf{u},\Mybf{\rho}_0,\Mybf{\rho}_q)\,
[\mathbb{Q}(u^*-is_0,\epsilon+\epsilon_0)]^\dagger\,\mathbb{Q}(u+is_0,\epsilon)=\\[3mm]
&=&\bar \chi(\Mybf{u},\Mybf{\rho}_0,\Mybf{\rho}_q)\,
\mathbb{Q}(u+i(1-s_0),\epsilon+\epsilon_0)\,[\mathbb{Q}(u^*-i(1-s_0),\epsilon)]^\dagger\,.
\ea
The normalization factors are given by the following expressions
\ba\label{QTchi}
\chi(\Mybf{u},\Mybf{\rho}_0,\Mybf{\rho}_q)&=&e^{i\pi(\epsilon_0+\epsilon_q)N}
\left[\frac{2^{-2iu}}{\pi}\right]^N\,c(\Mybf{\rho}_q)^{-N}\,\\
&&\times\left[
A(1+i\rho_q+i\rho_0-iu,\epsilon+\epsilon_q)\,A(1-i\rho_q-i\rho_0-iu,\epsilon+\epsilon_0)
\right]^{-N}\,,
\nonumber\\[2mm]
\bar
\chi(\Mybf{u},\Mybf{\rho}_0,\Mybf{\rho}_q)&=&(-1)^{2(\epsilon_0+\epsilon_q)N}
\,c(\Mybf{\rho}_q)^{-N}\,
r_{\mybf{\rho}_0\mybf{\rho}_q}(\Mybf{u})^N\,,
\ea
where $r_{\mybf{\rho}_0\mybf{\rho}_q}(\Mybf{u})$ is defined in
Eq.~(\ref{rn}). Thus an arbitrary transfer matrix~(\ref{TM}) can be
expressed  as the product of two Baxter~$\mathbb{Q}-$operators at
special values of the spectral parameter.

Using the diagrammatic representation,   
it is also straightforward to show
that the operators $\mathbb{Q}$ and $\mathbb{Q}^\dagger$ commute at arbitrary
values of the spectral parameters. Moreover, one can deduce the
following identity
\be\label{QQbar}
\left[(-1)^{2\epsilon_{v}}A(\Mybf{\alpha}_v)A(\Mybf{\beta}_v)\right]^{N}
\mathbb{Q}(\Mybf{u})\,[\mathbb{Q}(\Mybf{v^{*}})]^\dagger~=~
\left[(-1)^{2\epsilon_{u}}A(\Mybf{\alpha}_u)A(\Mybf{\beta}_u)\right]^{N}
\mathbb{Q}(\Mybf{v})\,[\mathbb{Q}(\Mybf{u^{*}})]^\dagger\,,
\ee
where $\Mybf{\alpha}_u$, $\Mybf{\beta}_u$ are defined in~(\ref{Y}).

Having put $\Mybf{\rho}_0=\Mybf{\rho}_q$ in Eq.~(\ref{T-QQ}) and
taking into account~(\ref{H}), one obtains the following expression for
the Hamiltonian $\CH_N$ in terms of the Baxter $\mathbb{Q}-$operator
\ba\label{H-Q}
\CH_N&=&i\left.\frac{d\phantom{u}}{du}\ln\mathbb{Q}(is_q+u,0)\right|_{u=0}~+~
i\left.\frac{d\phantom{u}}{du}\ln\mathbb{Q}(-is_q+u,\epsilon_q)^\dagger\right|_{u=0}~+~
\CE_N\nonumber\\[3mm]
&=&i\left.\frac{d\phantom{u}}{du}\ln u^N\mathbb{Q}(is_q^*+u,\epsilon_q)\right|_{u=0}~+~
i\left.\frac{d\phantom{u}}{du}\ln
u^N\mathbb{Q}(-is_q^*+u,0)^\dagger\right|_{u=0}~+~
\widebar\CE_N\,.
\ea
where the  additive constants $\CE_N$ and $\widebar \CE_N$ are
\be\label{CE}
\CE_N~=~N\left[\psi(2\epsilon_q+2i\rho_q)~+~\psi(2\epsilon_q-2i\rho_q)\right]\,,
\ \ \ \ \ \ \ \ \ \ \widebar \CE_N~=~2N\,\psi(1)\,.
\ee
Thus the Baxter
$\mathbb{Q}-$operator commutes  with the Hamiltonian $\CH_N$~(\ref{H}).
Evidently, the relations~(\ref{H-Q}), (\ref{P-Q}), (\ref{T-QQ}) hold
for the eigenvalues of the corresponding operators as well.
Thus, the knowledge of the eigenvalue of the Baxter
$\mathbb{Q}-$operator allows one to restore the eigenvalues of all other operators
in question. The remarkable property of the Baxter
$\mathbb{Q}-$operator is that its eigenvalues can be determined
without solving the eigenvalue problem. Indeed, since the Baxter
$\mathbb{Q}-$operator satisfies Eq.~(\ref{BE}), the same equation holds for its eigenvalues.
Therefore solving this equation in the
appropriate class of functions one can determine all eigenvalues of
the $\mathbb{Q}-$operator.
In order to be an eigenvalue of the Baxter $\mathbb{Q}-$operator
the solution of Eq.~(\ref{BE}) should satisfy the additional
conditions.
To find them
we shall study in the next section the
analytic properties of the eigenvalues of the $\mathbb{Q}-$operator as the function of $u$.

\subsection{Analytic structure and asymptotic}
\label{aa}
\begin{figure}[t]
\psfrag{1}[cc][cc]{\small{$1$}}
\psfrag{2}[cc][cc]{\small{$2$}}
\psfrag{3}[cc][cc]{\small{$3$}}
\psfrag{m1}[cc][cc]{\small{$-1$}}
\psfrag{m2}[cc][cc]{\small{$-2$}}
\psfrag{m3}[cc][cc]{\small{$-3$}}
\psfrag{0}[cc][cc]{}
\psfrag{r}[cc][cc]{$\rho_q$}
\psfrag{mr}[cc][cc]{$-\rho_q$}
\psfrag{im}[cc][cc]{$\Im\,(u)$}
\psfrag{re}[cc][cc]{$\Re\,(u)$}
\centerline{\epsfxsize9.0cm\epsfbox{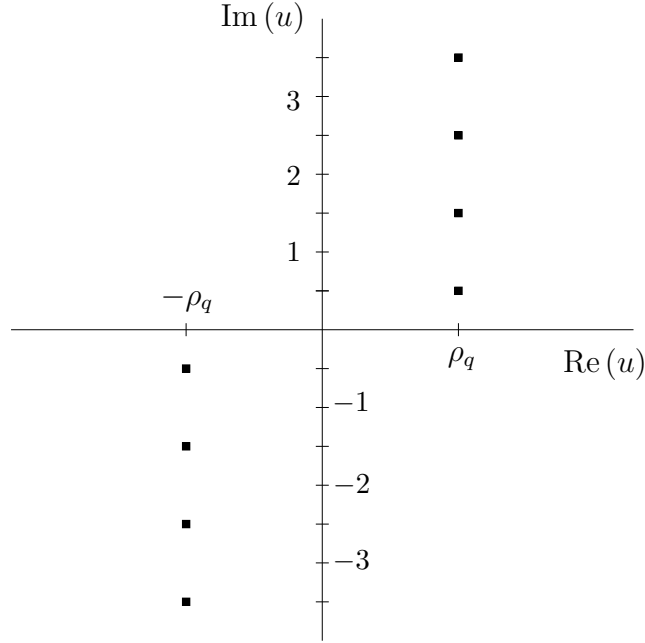}} \vspace*{0.5cm}
\caption{Distribution of the poles of the Baxter $\mathbb{Q}$-operator
  in the complex
$u$-plane for \mbox{$\epsilon=\epsilon_q=0$}.}\label{poles}
\end{figure}

We shall denote the eigenvalue of the Baxter $\mathbb{Q}-$operator corresponding to the
eigenfunction $\Psi_{q_2,\ldots,q_N}(\vec{x})$ as
$Q_{\mathbf{q}}(\Mybf{u})\equiv Q_{\mathbf{q}}(u,\epsilon)$
and study its properties as a
function of the complex variable $u$.
It follows directly from the definition~(\ref{Y}) and (\ref{QZ}) that the
Baxter $\mathbb{Q}-$operator is well defined as the integral operator
in the strip $-1/2<iu<1/2$. To continue it to the whole complex plane,
let us apply the Baxter operator, $\mathbb{Q}(\Mybf{u})$, to a test function $\Psi(\vec{x})$.
Then it can be seen that the resulting function
$\Phi_{\mybf{u}}(\vec{x})=[\mathbb{Q}(\Mybf{u})\Psi](\vec{x})$ as a
function of $u$ admits continuation to the whole complex plane except for
the points where the indices $\alpha$ and $\beta$ take positive
integer values, namely
$u_n^{-}=is_q-in,\ \  (\beta=n>0) $ and $u_n^{+}=-is_q+in \ \ (\alpha=n>0)$.
The structure of the singularities can be easily established using the
 representation of the Baxter $\mathbb{Q}-$operator given in
the rhs of Fig.~\ref{Q-diag}. Since the indices of the propagators
$\alpha'$ and $\beta'$ are not positive integers while $\alpha$ or
$\beta$ are, the corresponding integrals define a regular function of
$u$ at these points and all singularities are contained in the prefactor
$A(\Mybf{\alpha},\Mybf{\beta},\Mybf{\gamma})^N$. The straightforward
analysis shows that the Baxter $\mathbb{Q}-$operator has poles of the
order $N$ in the upper and lower half-planes in the following points
\begin{itemize}
\item{Upper half-plane}
\be\label{up}
\begin{tabular}{lll}
$\epsilon=\epsilon_q$, &$u_n^{+}=\rho_q+2in+\frac{i}{2},$&
  $n\geq 0$ \\[3mm]
$\epsilon\neq\epsilon_q$, & $u_n^{+}=\rho_q+2in,$&
  $n\geq 1$
\end{tabular}
\ee
\item Lower half-plane
\be\label{ud}
\begin{tabular}{lll}
$\epsilon=0$, &$u_n^{-}=-\rho_q-2in-\frac{i}{2},$&
  $n\geq 0$ \\[3mm]
$\epsilon=1/2$, & $u_n^{-}=-\rho_q-2in,$&
  $n\geq 1$
\end{tabular}
\ee
\end{itemize}
Thus the eigenvalue of the Baxter $\mathbb{Q}$-operator is a meromorphic
function of the variable $u$ with poles of the order $N$ at the
points $u_n^\pm(\epsilon)$, Eqs.~(\ref{up}) and  (\ref{ud}).
The positions of the poles in the $u$ plane are schematically shown in Fig.~\ref{poles}.
Note also that the equation in the second line of~(\ref{H-Q})
gives the energy as the ratio of the residues at the leading (of
order~$N$)  and the subleading (of order $N-1$) poles at $u_0^{\pm}=\pm (\rho_q+i/2)$.

Since the solutions of the Baxter equation~(\ref{BE}) are
determined up to multiplication
by the periodic function $f(\Mybf{u}+\mathbf{i})=f(\Mybf{u})$ one needs
additional information to fix this arbitrariness. This can be obtained by
the study of the asymptotic behaviour of the eigenvalues at large $u$.
First of all, let us notice that the transformation properties of the
eigenfunction under  $SL(2,\mathbb{R})$ rotation is determined by the eigenvalue
of the charge $q_2$ or the total Casimir operator,
$\vec{S}^2\equiv\left(\vec{S_1}+\ldots+\vec{S}_N\right)^2$ (see Eq.~(\ref{q2})).
Namely,
\mbox{$\vec{S}^2=-1/4-\rho^2$} if the eigenfunction transforms according the
irreducible representation of the principal series,
$T^{(\rho,\epsilon)}$,
and \mbox{$\vec{S}^2=h(h-1)$} if it transforms according to the discrete series
representation $\CD_h^\pm$. Thus depending on the value of the Casimir
operator, the eigenfunction can be represented as follows
\ba\label{eif-c}
\Psi_{\mathbf{q}}^{c}(x_1,\ldots,x_N)&=&\int_{-\infty}^\infty dx_0\,
\Pi_{q_3,\ldots,q_N}^{(\rho,\epsilon)}(x_1,\ldots,x_N|x_0)\,f^{(\rho,\epsilon)}(x_0)\,,\\
\label{eif-d}
\Psi_{\mathbf{q}}^d(x_1,\ldots,x_N)&=&
\int\CD_\pm w\,
\Pi_{q_3,\ldots,q_N}^{h,\pm}(x_1,\ldots,x_N|\bar w)\,\phi_h^\pm(w)\,,
\ea
where the integration measure in (\ref{eif-d}) is defined in Eq.~(\ref{mD}).
The projectors $\Pi^{(\rho,\epsilon)}$, $\Pi^{h,\pm}$ intertwine the
representations $(\otimes T^{\mybf{\rho_q}})^N$ and
$T^{(\rho,\epsilon)}$, $\CD_h^\pm$, respectively.
They have the following transformation properties
\ba\label{tran-P}
\Pi_{q_3,\ldots,q_N}^{(\rho,\epsilon)}(\vec{x}|x_0)&=&
\frac{\sigma_\epsilon(cx_0+d)}{|cx_0+d|^{2-2s}}
\prod_{k=1}^N
\frac{\sigma_{\epsilon_q}(cx_k+d)}{|cx_k+d|^{2s_q}}\,
\,\Pi_{q_3,\ldots,q_N}^{(\rho,\epsilon)}(\vec{x'}|x'_0)\,,\\
\Pi_{q_3,\ldots,q_N}^{h,\pm}(\vec{x}|\bar w)&=&
\frac{1}{(c\bar w+d)^{2h}}\prod_{k=1}^N
\frac{\sigma_{\epsilon_q}(cx_k+d)}{|cx_k+d|^{2s_q}}\,
\,\Pi_{q_3,\ldots,q_N}^{h,\pm}(\vec{x'}|\bar w')\,,
\ea
where $x'_k=(ax_k+b)/(cx_k+d)$ and $\bar w'=(a\bar w+b)/(c\bar w+d)$;
$a,b,c,d$ are real and  $ad-bc=1$. In particular, one finds that the
projectors are invariant under the simultaneous shift of all arguments by
a real number and are transformed under scale transformations as
\ba\label{Pshift}
\Pi_{q_3,\ldots,q_N}^{(\rho,\epsilon)}(\lambda x_1,\ldots,\lambda x_N|\lambda x_0)&=&
\lambda^{-1+s-Ns_q}\Pi_{q_3,\ldots,q_N}^{(\rho,\epsilon)}(x_1,\ldots,x_N|x_0)\,,\\[2mm]
\Pi_{q_3,\ldots,q_N}^{h.\pm}(\lambda x_1,\ldots,\lambda x_N|\lambda \bar w)&=&
\lambda^{-h-Ns_q}\Pi_{q_3,\ldots,q_N}^{h,\pm}(x_1,\ldots,x_N|\bar w)\,.
\ea
Applying the $\mathbb{Q}-$operator in the form (\ref{QZ}) to the
eigenfunction~(\ref{eif-c}), one finds that the leading contributions
at $u\to\infty$ comes from two integration regions over $\vec{y}$
\be\label{RI-II}
\mbox{\rm (I)}:\ \ \  |y_k|=\CO(u),\ \ \ \ \ \ \mbox{\rm (II)}:\ \ \ \ y_k-y_{k+1}=\CO(1/u)\,.
\ee
Next, let us notice that any function $f^{(\rho,\epsilon)}(x)$ can be
represented as the transformation of the function $f_0(x)=1$,
\mbox{$f^{(\rho,\epsilon)}(x)=\int Dg
\phi(g)T^{(\rho,\epsilon)}(g)f_0(x)$}, where the integral is taken over the group.
Then taking into account the invariance of the  Baxter
$\mathbb{Q}-$operator with respect to
the $SL(2,\mathbb{R})$ transformations and the properties of the projector
$\Pi_{q_3,\ldots,q_N}^{(\rho,\epsilon)}$~(\ref{tran-P}),
one can derive (see Ref.~\cite{DKM-I} for details)
\be\label{as-Qc}
Q_{\mathbf{q}}(u,\epsilon)\stackrel{u\to \infty}{=} \left(A_{I}\,u^{s-Ns_q}~+~A_{II}\,
u^{1-s-Ns_q}\right)\,\left[1+\CO(1/u)\right]\,.
\ee
The constants $A_{I}$ and $A_{II}$ depend on the integrals of
motion $q_k$, and we remind that $s=1/2+i\rho$, $s_q=1/2+\rho_q$.

In the case of the eigenfunctions of the discrete
spectrum~(\ref{eif-d}), a careful analysis shows that the
contributions coming from the regions $(I)$ and
$(II)$~(\ref{RI-II}) are of the same order
\be\label{as-Qd}
Q_{\mathbf{q}}(u,\epsilon)\stackrel{u\to \infty}{=} C\,u^{1-h-Ns_q}\,\left[1+\CO(1/u)\right]\,.
\ee
Let us also note that since the kernels of the operators
$\mathbb{Q}(u,0)$ and $\mathbb{Q}(u,1/2)$ coincide in the regions $(I)$
and $(II)$, the difference $Q_{\mathbf{q}}(u,0)-Q_{\mathbf{q}}(u,1/2)$
vanishes faster than any degree of $1/u$, i.e.
\be\label{diffQ}
Q_{\mathbf{q}}(u,0)-Q_{\mathbf{q}}(u,1/2)\stackrel{u\to \pm\infty}{=}\CO(e^{-\kappa |u|})\,,
\ee
where $\kappa$ is some  constant.

Thus  the solution of the Baxter equation~(\ref{BE}) corresponds to
some eigenvalue of the Baxter $\mathbb{Q}$-operator only if it has
the proper pole structure~(\ref{up}), (\ref{ud})  and proper
asymptotic~(\ref{as-Qc}), (\ref{as-Qd}) at $u\to\pm \infty$.

\section{Separation of Variables}
\label{SoV}
In this section we explicitly construct the representation of the Separated
Variables for the model in question. 
Namely, we shall obtain the following integral representation for the
eigenfunctions of the model
\be\label{SoV-rep}
\Psi_{\mathbf{q}}(\vec{x})~=~\frac{1}{2\pi}\int dp\int
\prod_{k=1}^{N-1}
\CD \Mybf{u}_k\,\mu_N(\vec{\Mybf{u}})\,
\Phi_{\mathbf{q}}(p,\Mybf{u}_1,\ldots,\Mybf{u}_{N-1})
U_{p,\mybf{u}_1,\ldots,\mybf{u}_{N-1}}(\vec{x})\,,
\ee
where
\be
\int\CD \Mybf{u}_k\equiv\frac1{(2\pi)^2}\int_{-\infty}^\infty du_k\, \sum_{\epsilon_k=0,1/2}\,,
\ee
$\mu_N(\vec{\Mybf{u}})$ is the Sklyanin measure (see Eq.~(\ref{measure})),
\ $\Phi_{\mathbf{q}}(p,\Mybf{u}_1,\ldots,\Mybf{u}_{N-1})$ is the
eigenfunction in the SoV representation~(\ref{Phi-Sov}) and
$U_{p,\mybf{u}_1,\ldots,\mybf{u}_{N-1}}(x_1,\ldots,x_N)$ is the
transition kernel to the SoV representation~(\ref{Up}).

In Sklyanin's approach~\cite{Sklyanin} the functions
$U_{p,\mybf{u}_1,\ldots,\mybf{u}_{N-1}}(x_1,\ldots,x_N)$ are identified
as the eigenfunctions of the operator $B_N(u)$~---~an off diagonal matrix
element of the monodromy matrix. Provided that the spectrum of the
operator $B_N(u)$ is non-degenerate, one can derive that the
eigenfunction in the SoV representation
($\Phi_{\mathbf{q}}(p,\Mybf{u}_1,\ldots,\Mybf{u}_{N-1})$) satisfies
the Baxter equation in each variable. In the case under consideration this
condition is not fulfilled. Moreover, the hermitian operator $B_N(u)$
admits different non-equivalent self-adjoint extensions and one has
to choose the correct one. Although these problems can be overcome, we
shall construct the transition kernel
$U_{p,\mybf{u}_1,\ldots,\mybf{u}_{N-1}}(x_1,\ldots,x_N)$ following
another approach. It was conjectured in Ref.~\cite{KS} that the
transition kernel to the SoV representation  can be related to the
kernel corresponding to the product of the Baxter
$\mathbb{Q}-$operators. The representation of such type is known now
for a number of spin chain models~\cite{DKM-I,DKM,open}. We suggest the
following ansatz for the transition kernel
\be\label{U}
U_{p,\mybf{u}_1,\ldots,\mybf{u}_{N-1}}(\vec{x})=c_N(p)
\int \prod_{k=1}^Ndy_k\, e^{ipy_1}\,
\left[\mathbb{Q}(\Mybf{u}_1)\ldots\mathbb{Q}(\Mybf{u}_{N-1})\right]
(x_1,\ldots,x_N|y_1,\ldots,y_N)
\ee
and show that so constructed kernel possesses
all necessary properties.
For convenience we put
$$
c_N(p)=|p|^{-N+1/2}\,c(\Mybf{\rho}_q)^{-N(N-1)/2}\,,
$$
where $c(\Mybf{\rho}_q)$ is defined in (\ref{cN}).

First of all let us notice that, as follows
from the properties of the Baxter $\mathbb{Q}$-operator, the
kernel~(\ref{U}) is a symmetric function of the
variables $\Mybf{u}_1,\ldots,\Mybf{u}_{N-1}$. Second, it has a
definite momentum \mbox{$i(S_{-}^{(1)}+\cdots+S_{-}^{(N)})
U_{p,\vec{\mybf{u}}}(\vec{x})=
pU_{p,\vec{\mybf{u}}}(\vec{x})$} and, at last, it satisfies the
Baxter equation~(\ref{BE}) in each variable $\Mybf{u}_k$.

\begin{figure}[t]
\psfrag{x1}[cc][cc]{$x_1$}
\psfrag{x2}[cc][cc]{$x_2$}
\psfrag{x3}[cc][cc]{$x_3$}
\psfrag{xN}[cc][cc]{$x_N$}
\psfrag{a1}[cc][cc]{\footnotesize{$\Mybf{\alpha}_1'$}}
\psfrag{a2}[cc][cc]{\footnotesize{$\Mybf{\alpha}_2'$}}
\psfrag{aN1}[cc][cc]{\footnotesize{$\Mybf{\alpha}_{N-1}'$}}
\psfrag{b1}[cc][cc]{\footnotesize{$\Mybf{\beta}_1'$}}
\psfrag{b2}[cc][cc]{\footnotesize{$\Mybf{\beta}_2'$}}
\psfrag{bN1}[cc][cc]{\footnotesize{$\Mybf{\beta}_{N-1}'$}}
\psfrag{g}[cc][cc]{\footnotesize{$\Mybf{\gamma}'$}}
\psfrag{dots}[cc][cc]{}
\psfrag{exp}[cc][cc]{$e^{ipy}$}
\centerline{\epsfxsize10.0cm\epsfbox{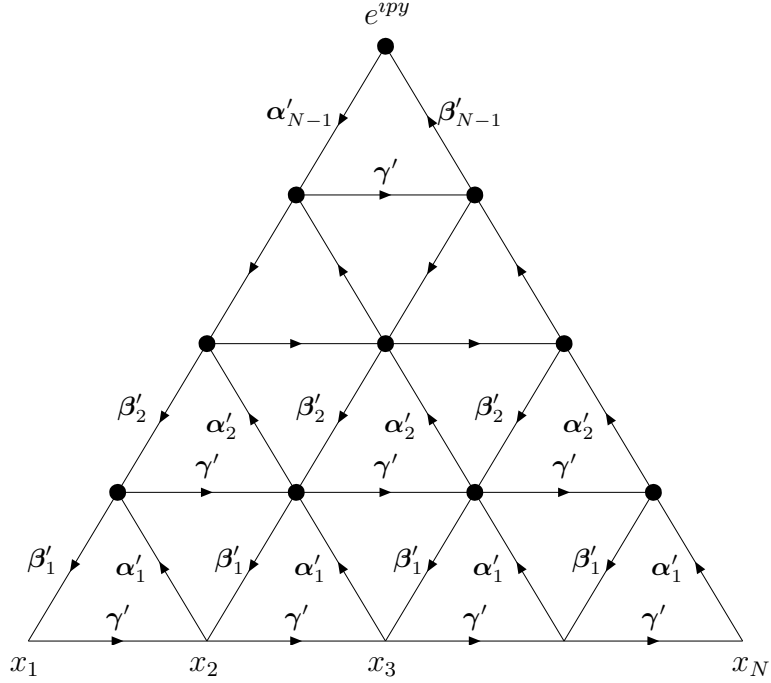}} \vspace*{0.5cm}
\caption{The graphical representation of the  transition
  kernel, Eq. (\ref{Up}).}
\label{UQ-diag}
\end{figure}

Further, we shall show
that for real $\Mybf{u}_k$, $k=1,\ldots,N-1$, the
functions $U_{p,\vec{\mybf{u}}}(\vec{x})$ are mutually orthogonal
\be\label{Ort}
\int\prod_{k=1}^N dx_k
\left(U_{q,\vec{\mybf{v}}}(\vec{x})\right)^*\,U_{p,\vec{\mybf{u}}}(\vec{x})~=~
2\pi\delta(p-q)\,\delta(\vec{\Mybf{u}}-\vec{\Mybf{v}})\,
\frac{\mu_N^{-1}(\vec{\Mybf{u}})}{(N-1)!}\,,
\ee
where
$$
\delta(\vec{\Mybf{u}}-\vec{\Mybf{v}})=\sum_S\delta(\Mybf{u}_1-\Mybf{v}_{i_1})\ldots
\delta(\Mybf{u}_{N-1}-\Mybf{v}_{i_{N-1}})\,,
$$
the sum goes  over all permutations and
$\delta(\Mybf{u}-\Mybf{v})=(2\pi)^2\delta_{\epsilon_u\epsilon_v}\delta(u-v)$.
To prove~(\ref{Ort}) it is convenient to represent the kernel~(\ref{U})
in another form. Namely, let us use the diagrammatic representation for
the Baxter $\mathbb{Q}$-operator shown in the rhs of Fig.~\ref{Q-diag}.
The diagram for the product of $N-1$ $\mathbb{Q}$-operators consists of
$N-1$ such rows. We shall assume that each triangle
in~Fig.~\ref{Q-diag} accompanied by the corresponding scalar factor
$\sqrt{\pi}A(\Mybf{\alpha},\Mybf{\beta},\Mybf{\gamma})$. According to
Eq.~(\ref{U}) one has to perform the integration over variables
$y_2,\ldots,y_N$ which correspond to the upper vertices of the $N-1$
triangles in the upper row. Since only two propagators are attached to
each vertex in question, one can use the chain relation,
Eq.~(\ref{chain}). The resulting propagators have the index
$\Mybf{\gamma}=(1-2s_q,\epsilon_q)$ and cancel the $N-1$ horizontal
lines in the upper row of the diagram. Thus after the integration over its
upper vertex, the triangle disappears producing the scalar factor
$c(\Mybf{\rho}_q)=\pi|A((1+2i\rho_q,\epsilon_q))|^2$. As a consequence
one obtains a new diagram in which $N-1$ triangles in the upper row
are eliminated and the additional  scalar factor
$c(\Mybf{\rho}_q)^{N-1}$ is acquired. Repeating this operation $N-1$ times one derives
that the kernel~(\ref{U}) is represented by the diagram shown in
Fig.~\ref{UQ-diag}
with the accompanying additional  factor $c(\Mybf{\rho}_q)^{N(N-1)/2}$.
Therefore the function~(\ref{U}) can be represented in the form
\be\label{Up}
U_{p,\mybf{u}_1,\ldots,\mybf{u}_{N-1}}(\vec{x})~=~|p|^{-N+1/2}\int dy\, e^{ipy}\,
\left[\Lambda_N(\Mybf{u}_1)\ldots\Lambda_2(\Mybf{u}_{N-1})\right]
(x_1,\ldots,x_N|\, y),
\ee
where  the
function $\Lambda_k(x_1,\ldots,x_k|y_1,\ldots,y_{k-1})(\Mybf{u}_{N+1-k})$
corresponds to the $k-1-$th row (from the top) of triangles in the
diagram shown in Fig.~\ref{UQ-diag}.
We remind here that each triangle is accompanied by the corresponding
scalar factor.

The representation~(\ref{Up}) is convenient for the calculation of the
scalar product~(\ref{Ort}). Using the representation~(\ref{Up}) for  both $U-$functions,
one finds that the scalar product~(\ref{Ort}) takes the form (up to some prefactor)
\be\label{UU}
\int dy'\,e^{-iqy'}\int dy\,e^{ipy}\left[
\Lambda_2(\Mybf{v}_1)^\dagger\ldots\Lambda_N(\Mybf{v}_{N-1})^\dagger\,
\Lambda_N(\Mybf{u}_{1})\ldots\Lambda_2(\Mybf{u}_{N-1})\right](y'\vert\, y)\,.
\ee
It can be easily checked that for real $\Mybf{u}$, $\Mybf{v}$, ($\Mybf{u}\neq\Mybf{v}$),
the $\Lambda-$operators satisfy the
following (exchange) relation
\be\label{Lambda-exchange}
\Lambda_k(\Mybf{v})^\dagger\,\Lambda_k(\Mybf{u})~=~c(\Mybf{\rho}_q)\,\varphi(\Mybf{u},\Mybf{v})\,
\Lambda_{k-1}(\Mybf{u})\,\Lambda_{k-1}(\Mybf{v})^\dagger\,
\ee
with
\be\label{vphi}
\varphi(\Mybf{u},\Mybf{v})~=~\frac{\pi\,(-1)^{2\epsilon_u+2\epsilon_v}}{
A(\Mybf{\beta}_v-\Mybf{\beta}_u)A(\Mybf{\alpha}_v-\Mybf{\alpha}_u)}\,.
\ee
Using this relation one obtains for (\ref{UU})
\ba\label{UU-1}
&&c(\Mybf{\rho}_q)^{(N-1)(N-2)/2}
\prod_{j=1}^{N-2}\prod_{k=1}^{N-1-j}\varphi(\Mybf{u}_k,\Mybf{v}_{N-j})
 \\
&&\ \ \ \ \ \ \times\int dy'\,e^{-iqy'}\int dy\,e^{ipy}
\left[\Lambda_2(\Mybf{v}_1)^\dagger\,
\Lambda_2(\Mybf{u}_{1})\,\ldots
\Lambda_2(\Mybf{v}_{N-1})^\dagger\Lambda_2(\Mybf{u}_{N-1})\right](y'\vert\, y)\,,\nonumber
\ea
where we suppose that $\Mybf{u}_k\neq\Mybf{v}_j$ if  $k\neq
j$. The calculation of the
convolution of the $\Lambda$ functions in the second line of
(\ref{UU-1}) (which
can be represented as a chain of the box diagrams) is relied on the
identity
\be\label{LL}
\int  dy\,e^{ipy}\left[\Lambda_2(\Mybf{v})^\dagger\Lambda_2(\Mybf{u})\right](x\vert\,y)~=~
c(\Mybf{\rho}_q) \frac{e^{ipx}}{|p|}\,(2\pi)^2 \,\delta_{\epsilon_u\epsilon_v}\,\delta({u}-{v})\,.
\ee
Collecting all factors together, one derives for (\ref{UU})
\be\label{UU-2}
\left[c(\Mybf{\rho}_q)^{N(N-1)/2}\prod_{m=1}^{N-2}\prod_{j=m+1}^{N-1}
\varphi(\Mybf{u}_m,\Mybf{u}_j)\right]
(2\pi)\delta(p-q)\prod_{k=1}^{N-1}\delta(\Mybf{u}_k-\Mybf{v}_k)\,,
\ee
where
$\delta(\Mybf{u}-\Mybf{v})=(2\pi)^2\delta_{\epsilon_u\epsilon_v}\delta(u-v)$.
Restoring  the symmetry with respect to the permutations of
variables $\Mybf{u}_1,\ldots,\Mybf{u}_{N-1}$
($\Mybf{v}_1,\ldots,\Mybf{v}_{N-1}$), which is broken
due to the imposed condition,
$\Mybf{u}_k\neq\Mybf{v}_j$ for  $k\neq j$, one arrives to the
expression in the rhs of Eq.~(\ref{Ort}) with
\be\label{measure}
\mu_N(\vec{\Mybf{u}})~=~\frac{1}{(N-1)!}
\left[c(\Mybf{\rho}_q)^{-N(N-1)/2}\prod_{m=1}^{N-2}\prod_{j=m+1}^{N-1}
\omega(\Mybf{u}_m,\Mybf{u}_j)\right]\,,
\ee
where
\be\label{omega}
\omega(\Mybf{u}_m,\Mybf{u}_j)=\varphi(\Mybf{u}_m,\Mybf{u}_j)^{-1}=
\frac1{\pi}\frac{u_m-u_j}{2}\left[\tanh\pi\left(\frac{u_m-u_j}{2}\right)\right]^{1-
4(\epsilon_m+\epsilon_j)}\,.
\ee
We remind here that $(\epsilon_m+\epsilon_j)$ is equal to zero if
$\epsilon_m=\epsilon_j$ and to $1/2$ otherwise.

The measure $\mu_N(\vec{\Mybf{u}})$ is a regular nonnegative function for real
$\Mybf{u}_k$. When one of the separated variables $u_k$ goes to
infinity, it grows as
\be\label{asm}
\mu_N(\vec{\Mybf{u}})\stackrel{u_k\to
  \pm\infty}{\sim}~|u_k|^{N-2}\left(1+\CO(1/|u|)\right)\,.
\ee
Under the analytic continuation to the complex plane, the measure~(\ref{measure}) becomes
a meromorphic function of the variables $u_k$.  One can also verify
that it satisfies the functional relation
\be\label{f-asm}
\frac{\mu_N(\Mybf{u}_1,\ldots,\Mybf{u}_k+\mathbf{i},\ldots,\Mybf{u}_{N-1})
}{\mu_N(\Mybf{u}_1,\ldots,\Mybf{u}_k,\ldots,\Mybf{u}_{N-1})}~=~
\prod_{j\neq k}\frac{u_k-u_j+i}{u_k-u_j}\,.
\ee

To determine the wave function in the SoV representation,
$\Phi_{q_2,\ldots,q_N}(p,\Mybf{u}_1,\ldots,\Mybf{u}_{N-1})$, one
calculates the scalar product of the eigenfunction in the coordinate
representation,
$\Psi_{\mathbf{q}}(\vec{x})
$, (Eq.~(\ref{SoV-rep})) with
$U_{p,\mybf{u}_1,\ldots,\mybf{u}_{N-1}}(\vec{x})$, (Eq.~(\ref{U})).
Using Eq.~(\ref{Ort}) one finds
\be\label{Phi-Sov}
\Phi_{\mathbf{q}}(p,\Mybf{u}_1,\ldots,\Mybf{u}_{N-1})~=~
c_\Psi(p)\, Q_{\mathbf{q}}(\Mybf{u}_1)^*\ldots Q_{\mathbf{q}}(\Mybf{u}_{N-1})^*\,,
\ee
where
\be\label{cpsi}
c_\Psi(p)~=~c_N(p)\int dx_1\ldots dx_N\,e^{-ipx_1}\Psi_{\mathbf{q}}(x_1,\ldots,x_N)\,.
\ee
Assuming that the basis formed by the functions
$U_{p,\mybf{u}_1,\ldots,\mybf{u}_{N-1}}(\vec{x})$ is complete, one
derives the orthogonality condition for the wave functions in the SoV
representations
\be\label{SoV-ort}
\int
\prod_{k=1}^{N-1}
\CD \Mybf{u}_k\,\mu_N(\vec{\Mybf{u}})\, \prod_{k=1}^{N-1}
Q_{\mathbf{q}'}(\Mybf{u}_k)  Q_{\mathbf{q}}(\Mybf{u}_k)^*\sim \delta_{{\mathbf{q}}\mathbf{q}'}\,,
\ee
where we took into account that the integral over $p$ is factorized.

\section{Special case: $N=2$}
\label{N2}

In the case of the two-sites spin chain, the eigenvalues of the Baxter
$\mathbb{Q}-$operator can be obtained in a closed form. Indeed, for the $N=2$
spin chain, the eigenfunctions are fixed by the group properties,
Eqs.~(\ref{Proc}), (\ref{Prod}). Henceforth, for brevity,  we shall
restrict ourselves to the case of the representation of the positive
parity, i.e. we shall imply that $\epsilon_q=0$. Applying the Baxter
$\mathbb{Q}-$operator to the  eigenfunctions and going through the
calculations, one finds the representation of the Baxter function in the
form of a one-dimensional integral.
We shall denote the eigenvalue of the Baxter operator corresponding to
the eigenfunction of the continuous series
$(\Pi^{\mybf{\rho},\varepsilon}_{\mybf{\rho}_q\mybf{\rho}_q}
(x,y_1,y_2))^*$, (see Eq.~(\ref{Proc})) by
 $Q_\rho^\varepsilon(u,\epsilon)$~\footnote{Since we assume that $\Mybf{\rho_q}=(\rho_q,0)$,
the irreducible representation of the positive parity  ($\epsilon=0$)
appears in the tensor product decomposition, therefore
$\Mybf{\rho}=(\rho,0)$. We  remind also that the parameter $\varepsilon=0, 1/2$ marks
two different eigenfunctions having the same value of the two-particle Casimir
operator.}.
The eigenvalues of the Baxter operator corresponding to the
eigenfunctions of the discrete series
are the same, $\mathbb{Q}(u,\epsilon)(\Pi^{h,\pm}_{\mybf{\rho}_1\mybf{\rho}_2}(w,x_1,x_2))^*
= Q_h(u,\epsilon)(\Pi^{h,\pm}_{\mybf{\rho}_1\mybf{\rho}_2}(w,x_1,x_2))^*$.
Up to
unessential here $\Mybf{u}-$independent factors the eigenvalues of the
Baxter operator are given by the following expressions
\ba\label{Qdh}
Q_h(u,\epsilon)&=&\widehat Q_h(u,\epsilon)+(-1)^h\,\widehat Q_h(-u,\epsilon)\,,\\[2mm]
\label{Qch}
Q_{\rho}^\varepsilon(u,\epsilon)&=&
\widehat Q_\rho^{\varepsilon}(u,\epsilon)+(-1)^{2\varepsilon}
\widehat Q_\rho^{\varepsilon}(-u,\epsilon)\,,
\ea
where
\ba\label{QhQ}
\widehat Q_h(u,\epsilon)&=&
\int_0^1{d\tau}\tau^{iu-1}\,
\left[q_h(\tau)+(-1)^{2\epsilon}\,q_h(-\tau)\right]\,,\\[2mm]\label{QeQ}
\widehat Q_\rho^{\varepsilon}(u,\epsilon)&=&
\int_{0}^1
{d\tau}\,\tau^{iu-1}\left[
q_{\rho}^{\varepsilon}(\tau)+(-1)^{2\epsilon}q_{\rho}^{\varepsilon}(-\tau)\right]\,.
\ea
The functions $q_h$ and $q_\rho^\varepsilon$ can be expressed in terms
of the Legendre functions of the second kind as follows
\ba\label{qh}
q_h(\tau)&=&\left(\frac{|\tau|}{(1-\tau)^{2}}\right)^{1-s_q}\,
\mathbf{Q}_{h-1}\left(\frac{1+\tau}{1-\tau}\right)\,,\\[2mm]
\label{qc}
q_{\rho}^{\varepsilon}(\tau)&=&\left(\frac{|\tau|}{(1-\tau)^{2}}\right)^{1-s_q}
\,\left[
\lambda(\rho,\varepsilon)\,
\mathbf{Q}_{-1/2+i\rho}\left(\frac{1+\tau}{1-\tau}\right)+
\lambda(-\rho,\varepsilon)\,
\mathbf{Q}_{-1/2-i\rho}\left(\frac{1+\tau}{1-\tau}\right)\right]\,,
\ea
where
\be\label{zt}
\lambda(\rho,\varepsilon)=1+(-1)^{2\varepsilon}\frac{i}{\sinh\pi\rho}\,.
\ee
The argument of the Legendre function entering the integrals~(\ref{QhQ}), (\ref{QeQ})
varies in the range $[-1,\infty]$.  We remind here that for real $x$, $-1<x<1$, the
Legendre function is defined by the relation
\mbox{$\mathbf{Q}_\nu(x)=\frac12(\mathbf{Q}_\nu(x+i\epsilon)+(\mathbf{Q}_\nu(x-i\epsilon))$.}

The factors $(-1)^h,\ (-1)^{2\varepsilon}$ entering Eqs.~(\ref{Qdh}), (\ref{Qch})
determine the parity of the eigenfunctions with
respect to the permutation of the arguments.
Therefore, one concludes that the eigenvalue of the Baxter
operator is even (odd) function of $u$ for the eigenfunction of
positive (negative) parity.

Further, it is evident that the function $\widehat Q(u)$ ($\widehat Q(-u)$) has poles in
the upper (lower) half-plane. The poles of $\widehat Q(u)$ occur at the
points where the
integrals~(\ref{QhQ}) start to diverge at $\tau\to 0$.
Noticing that the integrand in (\ref{QhQ}), (\ref{QeQ})  is
an even ($\epsilon=0$) or odd ($\epsilon=1/2$)  function of
$\tau$ and  taking into account the properties of Legendre
functions, one concludes that
 the function
$Q(u,\epsilon)$ has poles of the second order located at the points~(\ref{up}), (\ref{ud}).
It is also straightforward to see that the asymptotic of the Baxter
functions~(\ref{Qdh}), (\ref{Qch}) at $u\to\infty$ is in
agreement with Eqs.~(\ref{as-Qc}), (\ref{as-Qd}). Calculating the
difference $Q(u,0)-Q(u,1/2)$ one finds that it can be represented as
\be\label{dQ}
\sim\int_0^\infty{d\tau}\tau^{iu-1}
q(-\tau)\,,
\ee
where $q(\tau)$ is given by (\ref{qh}) or (\ref{qc}). Since the
function $q(\tau)$ is analytic near $\tau=-1$ we reproduce (\ref{diffQ}).

At last, to find the energies of the corresponding eigenstate
it is necessary  to calculate the expansions of the Baxter functions near the first
poles,
$u=\pm i(1-s_q)$ (see. Eq.~(\ref{H-Q})).
Taking into account that for $\tau\to 0$
\be\label{Qto0}
\mathbf{Q}_{\nu}\left(\frac{1+\tau}{1-\tau}\right)=-\frac12\log|\tau|+
\psi(1)-\psi(1+\nu)+\CO(\tau)
\ee
one finds, e.g. for $\widehat Q_\rho^{\varepsilon}(-i(1-s_q)+u,0)$,
\ba\label{expQ}
&&\widehat Q_\rho^{\varepsilon}(-i(1-s_q)+u,0)=2
\int_{0}^1
\frac{d\tau}{\tau^{1-iu}}\left[-\log\tau +\right.\\[3mm]
&&\ \ \ \ \ \ \left.\lambda(\rho,\varepsilon)[\psi(1)-\psi(1/2+i\rho)]
+\lambda(-\rho,\varepsilon)[\psi(1)-\psi(1/2-i\rho)]\right]+\CO(u)\,.
\nonumber
\ea
Then using Eq.~(\ref{H-Q})
and taking into account that $\CH_2=2H_{12}$, one  reproduces the expressions~(\ref{ei-H}).

\section{Summary}
\label{Con}
In this paper we studied the spin chain model with the
$SL(2,\mathbb{R})$ symmetry group. The Hilbert space attached to each
site is
$L^2(\mathbb{R})$ and the symmetry transformations are realized by the
operators of the unitary principal series continuous representation of
the  $SL(2,\mathbb{R})$ group. To define the model we constructed
the  $SL(2,\mathbb{R})$ invariant solution of the Yang-Baxter equation
-- the $\CR-$operator which acts on the tensor product of two
$L^2(\mathbb{R})$ spaces.  The Hamiltonian of the model is
defined as the derivative of the fundamental transfer matrix
and is given by the sum of the
two-particle Hamiltonians. The pair-wise Hamiltonians have both
the discrete and continuous spectrum,~Eq.~(\ref{ei-H}), which reflects
the pattern of the  decomposition of the tensor product of two
representations of the principal continuous series into irreducible components.
The eigenstates belonging to the continuous spectrum are specified  by
the value of the $sl(2)$ spin, $s=1/2+i\rho$, and parity with respect
to the permutation of the coordinates $x_1$
and $x_2$. The energy gap between the eigenstates with the same value
of the spin and different
parity is maximal for $\rho=0$ and exponentially decreases with $\rho$.

To solve  the model we applied
the method of the Baxter $\mathbb{Q}-$operator~\cite{Baxter} and the Separation
of Variables~\cite{Sklyanin}.
The standard ABA
approach~\cite{ABA} is not applicable for the model in question
due to the absence of the lowest
weight vector in the Hilbert space of the model.
Having realized the Baxter operator as an integral operator we
resolved the defining equations and obtained the kernel in the
explicit form. It allowed to determine the properties of the
eigenvalues of the Baxter operator as functions of the spectral
parameter. Then the eigenvalues of the Baxter operator can be obtained
as the solutions of the Baxter equation in the certain class of functions.
The eigenvalue of the Baxter operator encodes all information about the
corresponding eigenstate.  We have shown that the Hamiltonian  of the model
can  be obtained  as a derivative of the Baxter operator at special points.
Moreover, the arbitrary transfer matrix factorizes into the
product of two Baxter $\mathbb{Q}$ operators at certain values of the
spectral parameters. Analoguous results have been obtained for the
noncompact $SL(2,\mathbb{R})$ (discrete series)  and
$SL(2,\mathbb{C})$ (continuous series) spin chains, see Refs.~\cite{SD,DKM-I}.

We have constructed the representation of the separated variables for
the model in question. The kernel of the unitary operator, which maps
the eigenfunction to the SoV representation, has been obtained in an explicit form.
It factorizes into the product of $N-1$ ($N$ is the
number of sites) operators each depending on one separated
variable only.  The kernel of the transition operator can be
visualized as a Feynman diagram with a specific pyramidal form.
This form of the kernel, first obtained for the $SL(2,\mathbb{C})$
spin chain~\cite{DKM-I}, is a general feature of all noncompact $SL(2)$ spin
magnets~\cite{DKM,open}. Using the diagram technique we
calculated the scalar product of the transition kernels and determined
the Sklyanin's integration measure. We have shown that the wavefunction in
the separated variables is given by the product of the eigenvalues of
the (conjugated) Baxter operator.
Therefore the  knowledge of the eigenvalue of the Baxter operator allows  to
restore the eigenfunction.

\section*{Acknowledgements}

We are grateful to S.~Derkachov and G.~Korchemsky for helpful  discussions. This work was
supported in part by the grant 03-01-00837 of the Russian Foundation for
Fundamental Research, the Sofya
Kovalevskaya programme of Alexander von Humboldt Foundation (A.M.) and
the Graduiertenkolleg 841 of the DFG (M.K.).


\setcounter{equation}{0}
\appendix
\renewcommand{\theequation}{\Alph{section}.\arabic{equation}}
\setcounter{table}{0}
\renewcommand{\thetable}{\Alph{table}}
\section{Appendix: Tensor product decomposition}
\label{AppA}

In this Appendix we collect  the necessary formulae  concerning
the decomposition of
the tensor product
of the representations of the principal continuous series.
The tensor product of two
representations of the principal continuous series
is decomposed into the direct integral over the representations of the
same type and into the direct sum  over the unitary representations of the
discrete series~\cite{Pk}.
We remind that the
representations of the discrete series ${\cal D}^{\pm}_h $
are labelled by the (half)integer number,~$h$, and can be
realized by the unitary operators $D^+_h(g)$ ($D^-_h(g)$)
\be\label{trd}
\left[D^{\pm}_h(g)\Psi_\pm\right](w)~=~(cw+d)^{-2h}\,\Psi_\pm\left(\frac{aw+b}{cw+d}\right),
\ee
acting on the Hilbert
space ${\cal H}^\pm_h$, respectively. The latter are defined as the
space of the functions analytic in upper (${\cal H}^+_h$),  lower (${\cal H}^-_h$)
half-plane with the scalar product defined by the following expression~\cite{Gelfand}
\be\label{scd}
\vev{\Psi_{\pm}|\Phi_\pm}~=\int \CD_{\pm} w \,
\overline{\Psi_\pm(w)}\,\Phi_\pm(w)\,,
\ee
where
\be\label{mD}
\CD_{\pm} w=\frac{2h-1}{\pi} \theta(\pm \Im(w)) (\pm \Im(w))^{2h-2} dxdy
\ee
 and $w=x+iy$.

The operators separating the irreducible components in the
tensor product $T^{\mybf{\rho}_1}\otimes T^{\mybf{\rho}_2}$ are the following:
the projectors to the discrete series, $\Pi^{h,\pm}_{\mybf{\rho}_1\mybf{\rho}_2}$, which map
 $T^{\mybf{\rho}_1}\otimes T^{\mybf{\rho}_2}\to D^{\pm}_h$, and the projectors
$\Pi^{\mybf{\rho},\varepsilon}_{\mybf{\rho}_1\mybf{\rho}_2}$ to the principal
continuous series. The projecting operators can be realized as
integral operators.
The irreducible representation $T^{\mybf{\rho}}$,
($\Mybf{\rho}=(\rho,\epsilon)$, $\rho>0$, $\epsilon=\epsilon_1+\epsilon_2$)
enters into the tensor product  $T^{\mybf{\rho}_1}\otimes
T^{\mybf{\rho}_2}$ with double multiplicity.
Therefore one can construct two projectors,
 $\Pi^{\mybf{\rho},\varepsilon}_{\mybf{\rho}_1\mybf{\rho}_2}$,
where the parameter $\varepsilon$ takes two values $0$ and $1/2$ and
marks these  equivalent representations.
The integral kernel of the projector  is given by the product of three propagators
\be\label{Proc}
\Pi^{\mybf{\rho},\varepsilon}_{\mybf{\rho}_1\mybf{\rho}_2}(x,y_1,y_2)
~=~
D_{\mybf{\alpha}_3}(y_2-x) D_{\mybf{\alpha}_2}(x-y_1)
D_{\mybf{\alpha}_1}(y_1-y_2)\,,
\ee
where the indices are defined as follows
\ba\label{Pcind}
\Mybf{\alpha_1}&=&(\frac12-i(\rho+\rho_1+\rho_2),\varepsilon+\epsilon_1),\nonumber \\
\Mybf{\alpha_2}&=&(\frac12+i(\rho+\rho_2-\rho_1),\varepsilon),\\
\Mybf{\alpha_3}&=&(\frac12+i(\rho+\rho_1-\rho_2),\varepsilon+\epsilon_1+\epsilon_2)\,.\nonumber
\ea
Then the function $\Phi^{\mybf{\rho},\varepsilon}$ defined as
\be\label{proj-c}
\Phi^{\mybf{\rho},\varepsilon}(x)~=~
\int dy_1dy_2\,\Pi^{\mybf{\rho},\varepsilon}_{\mybf{\rho}_1\mybf{\rho}_2}(x,y_1,y_2)\,
\Psi(y_1,y_2),
\ee
transforms according to the representation  $T^{\mybf{\rho}}$. Next,
using the techniques for the calculation of the $SL(2,\mathbb{R})$-integrals introduced in Sec.~\ref{R-operator}, it is straightforward to check
the following
orthogonality condition
\be\label{ortP}
\int dy_1\,dy_2
\Pi^{\mybf{\rho},\varepsilon}_{\mybf{\rho}_1\mybf{\rho}_2}(x,y_1,y_2)\,
\left(\Pi^{\mybf{\rho'},\varepsilon'}_{\mybf{\rho}_1\mybf{\rho}_2}(x',y_1,y_2)
\right)^*~=~{\cal N}(\Mybf{\rho})\,\delta_{\varepsilon\varepsilon'}\,\delta(x-x')\,
\delta(\rho-\rho'),
\ee
where
\be\label{ncP}
{\cal N}(\Mybf{\rho})~=~(2\pi)^2\rho^{-1}\coth^{1-4\epsilon}(\pi\rho)\,.
\ee
\vskip 0.5cm

The conformal spin $h$ of the representations of the discrete series
$D_h^\pm$ entering the decomposition of the  tensor product $T^{\mybf{\rho}_1}\otimes
T^{\mybf{\rho}_2}$ is integer for $\epsilon_1=\epsilon_2$, and
half-integer otherwise. The projecting operators onto irreducible
components are
\ba\label{Prod}
\Pi^{h,\pm}_{\mybf{\rho}_1\mybf{\rho}_2}(w,x_1,x_2)&=&
\frac{(x_2-x_1)^{h+\epsilon_1+\epsilon_2}}{
( x_1-w)^{h+\epsilon_2-\epsilon_1}
( x_2-w)^{h+\epsilon_1-\epsilon_2}}\,\,
|x_1-x_2|^{i(\rho_1+\rho_2)-1-\epsilon_1-\epsilon_2}\,\\[2mm]
&\times&\left[\frac{x_1-w}{ x_2-w}(x_1-x_2)\right]^{[i(\rho_1-\rho_2)-\epsilon_1+\epsilon_2]/2}\,
\left[\frac{ x_2-w}{
    x_1-w}(x_2-x_1)\right]^{[i(\rho_2-\rho_1)-\epsilon_2+\epsilon_1]/2}\nonumber\,,
\ea
where the $w$ lies in the upper half-plane for the $''+''$ projector,
and in the lower for $''-''$ projectors. Note that the expressions in
the square brackets in~(\ref{Prod}) are the single--valued functions of
$w$ in the upper or lower half-plane. Thus the functions $\Phi^{h,\pm}$
\be\label{proj-d}
\Phi^{h,\pm}(x)~=~
\int dy_1dy_2\,\Pi^{h,\pm}_{\mybf{\rho}_1\mybf{\rho}_2}(w,y_1,y_2)\,\Psi(y_1,y_2)
\ee
transform according to the representations $\CD_h^\pm$. The normalization
condition reads
\be
\int dx_1dx_2
\left(\Pi^{h,\pm}_{\mybf{\rho}_1\mybf{\rho}_2}(w,y_1,y_2)\right)^*
\,\Pi^{h',\pm}_{\mybf{\rho}_1\mybf{\rho}_2}(z,y_1,y_2)~=~
c(h)\,\delta_{hh'}\,\mathbb{K}(z,w),
\ee
where the $\mathbb{K}(z,w)=e^{i\pi h} (z-\bar w)^{-2h}$ is the reproducing kernel
(unit operator on $\CH_h^\pm $) and
\be\label{ch}
c(h)=(2\pi)^2\frac{\Gamma(2h-1)}{\Gamma^2(h)}\,.
\ee

Then any function $\Psi(x_1,x_2)\in L^2(\mathbb{R})\otimes L^2(\mathbb{R})$
can be decomposed as
\ba\label{decomposition}
\Psi(x_1,x_2)&=&\sum_{h=1+(\epsilon_1+\epsilon_2)/2}^\infty c^{-1}(h)\int \CD_{\pm} w
\left(\Pi^{h,\pm}_{\mybf{\rho}_1\mybf{\rho}_2}(w,y_1,y_2)\right)^*
\Phi^{h,\pm}(w)~+~\nonumber\\
&&
\sum_{\varepsilon=0,1/2}\,\int_0^\infty d\rho\,
\CN^{-1}(\Mybf{\rho})\,\int_{-\infty}^{\infty} dx
\,\left(\Pi^{\mybf{\rho},\varepsilon}_{\mybf{\rho}_1\mybf{\rho}_2}(x,x_1,x_2)\right)^*
\,\Phi^{\mybf{\rho},\varepsilon}(x)\,,
\ea
where the functions $\Phi^{h,\pm} $ and
$\Phi^{\mybf{\rho},\varepsilon} $ are defined in
Eqs.~(\ref{proj-d}), (\ref{proj-c}), respectively.

Note, that if $\Mybf{\rho}_1=\Mybf{\rho}_2$, the projectors have
definite parity with respect to the permutation of the coordinates  $x_1$ and $x_2$,
$$
\Pi^{\mybf{\rho},\varepsilon}_{\mybf{\rho}_1\mybf{\rho}_1}
(x,x_1,x_2)~=~(-1)^{2\varepsilon+2\epsilon_1}\,
\Pi^{\mybf{\rho},\varepsilon}_{\mybf{\rho}_1\mybf{\rho}_1}(x,x_2,x_1)
$$
and
$$
\Pi^{h,\pm}_{\mybf{\rho}_1\mybf{\rho}_1}(w,x_1,x_2)~=~(-1)^{h+2\epsilon_1}\,
\Pi^{h,\pm}_{\mybf{\rho}_1\mybf{\rho}_1}(w,x_2,x_1)\,.
$$


\end{document}